\pdfoutput=1
\documentclass[aps,prd,amsmath,floats,floatfix, twocolumn,
superscriptaddress,nofootinbib,showpacs,longbibliography]{revtex4-1}
\usepackage[T1]{fontenc}
\usepackage[utf8]{inputenc}
\usepackage{txfonts}
\usepackage{verbatim}
\usepackage[dvipsnames, usenames]{xcolor}
\definecolor{linkcolor}{rgb}{0.0,0.3,0.5}
\usepackage[hypertexnames=false, unicode, colorlinks=true, linkcolor=linkcolor,
citecolor=linkcolor, filecolor=linkcolor,urlcolor=linkcolor,
pdfusetitle]{hyperref}
\usepackage[all]{hypcap}
\usepackage{graphicx}
\usepackage{xspace}
\usepackage{amssymb}
\usepackage{microtype}
\usepackage{subfigure}
\usepackage{url}
\usepackage[english]{babel}
\usepackage{blindtext}
\usepackage[normalem]{ulem} 
\usepackage{bm} 
\usepackage{mathrsfs}
\usepackage{xspace} 
\usepackage{fontawesome}

\DeclareMathAlphabet{\mathpzc}{OT1}{pzc}{m}{it}

\usepackage[nomessages]{fp}

\newcommand{\sk}[1]{}

\begin{document}
\title{\texttt{gwharmone:} first data-driven surrogate \\for eccentric harmonics in binary black hole merger waveforms}
\newcommand{\KITP}{\affiliation{Kavli Institute for Theoretical Physics, University of California Santa Barbara, Kohn Hall, Lagoon Rd, Santa Barbara, CA 93106}} 
\newcommand{\TAPIR}{\affiliation{Theoretical AstroPhysics Including Relativity and Cosmology, California Institute of Technology, Pasadena, California, USA}}
\author{Tousif Islam}
\email{tousifislam@ucsb.edu}
\KITP
\TAPIR
\author{Tejaswi Venumadhav}
\affiliation{\mbox{Department of Physics, University of California at Santa Barbara, Santa Barbara, CA 93106, USA}}
\affiliation{\mbox{International Centre for Theoretical Sciences, Tata Institute of Fundamental Research, Bangalore 560089, India}}
\author{Ajit Kumar Mehta}
\affiliation{\mbox{Department of Physics, University of California at Santa Barbara, Santa Barbara, CA 93106, USA}}
\author{Isha Anantpurkar}
\affiliation{\mbox{Department of Physics, University of California at Santa Barbara, Santa Barbara, CA 93106, USA}}
\author{Digvijay Wadekar}
\affiliation{\mbox{Department of Physics and Astronomy, Johns Hopkins University,
3400 N. Charles Street, Baltimore, Maryland, 21218, USA}}
\affiliation{\mbox{School of Natural Sciences, Institute for Advanced Study, 1 Einstein Drive, Princeton, NJ 08540, USA}}
\author{Javier Roulet}
\affiliation{\mbox{TAPIR, Walter Burke Institute for Theoretical Physics, California Institute of Technology, Pasadena, CA 91125, USA}}
\author{Jonathan Mushkin}
\affiliation{\mbox{Department of Particle Physics \& Astrophysics, Weizmann Institute of Science, Rehovot 76100, Israel}}
\author{Barak Zackay}
\affiliation{\mbox{Department of Particle Physics \& Astrophysics, Weizmann Institute of Science, Rehovot 76100, Israel}}
\author{Matias Zaldarriaga}
\affiliation{\mbox{School of Natural Sciences, Institute for Advanced Study, 1 Einstein Drive, Princeton, NJ 08540, USA}}

\hypersetup{pdfauthor={Islam et al.}}
\date{\today}

\begin{abstract}
We present \texttt{gwharmone}, the first data-driven surrogate model for eccentric harmonics (as well as the full radiation content) of the dominant quadrupolar mode in eccentric, non-spinning binary black hole mergers.  
Our model is trained on 173 waveforms, each $100,000M$ long (where $M$ is the total mass), generated for mass ratios $q \in [1,3.5]$ and eccentricities $e_{\rm ref} \in [0,0.2]$ (at the start of the waveform). The eccentric harmonics are extracted from the effective-one-body waveforms using the \texttt{gwMiner} package.  
We apply a singular value decomposition (SVD) to obtain a set of reduced basis vectors, necessary to construct a lower-dimensional representation of data, and use Gaussian Process Regression (GPR) to interpolate SVD coefficients across parameter space, allowing for prediction at new parameter points. 
The model includes the effect of mean anomaly, its evaluation cost is only $\sim 0.1$ second and it achieves an average time-domain (validation) error of $\sim 10^{-3}$ and frequency-domain (validation) mismatches below $10^{-2}$ for advanced LIGO sensitivity. Our model can therefore be useful in efficient searches and parameter estimation of eccentric mergers. \texttt{gwharmone} will be publicly available through the \texttt{gwModels} package. \href{https://github.com/tousifislam/gwModels}{\faGithub}
\end{abstract}
\maketitle

\noindent {\textbf{\textit{Introduction}}.}
Eccentric binary black hole (BBH) mergers are among the most anticipated sources of gravitational waves (GWs), given their rarity in observational data so far~\cite{Harry:2010zz,VIRGO:2014yos,KAGRA:2020tym,LIGOScientific:2018mvr,LIGOScientific:2020ibl,LIGOScientific:2021usb,LIGOScientific:2021djp}. These systems have garnered significant attention due to the complex features they introduce into waveforms and their potential to probe dense stellar clusters and galactic nuclei, where they are most likely to form. More importantly, measuring eccentricity can help distinguish different formation channels of BBH mergers~\cite{Rodriguez:2017pec,Rodriguez:2018pss,Samsing:2017xmd,Zevin:2018kzq,Zevin:2021rtf,Samsing:2020tda}. However, achieving this requires faithful and computationally efficient waveform models, which are essential for developing robust detection pipelines and parameter estimation frameworks. Recently, considerable efforts have been made to perform large-scale eccentric numerical relativity (NR) simulations~\cite{Mroue:2010re, Healy:2017zqj,Buonanno:2006ui,Husa:2007rh,Ramos-Buades:2018azo,Ramos-Buades:2019uvh,Purrer:2012wy,Bonino:2024xrv,Ramos-Buades:2022lgf}, investigate eccentricity effects within the post-Newtonian (PN) calculations~\cite{Arun:2009mc,Tanay:2016zog,Paul:2022xfy,Henry:2023tka}, incorporate eccentricity into waveform models~\cite{Tiwari:2019jtz, Huerta:2014eca, Moore:2016qxz, Damour:2004bz, Konigsdorffer:2006zt, Memmesheimer:2004cv,Hinder:2017sxy, Cho:2021oai,Chattaraj:2022tay,Hinderer:2017jcs,Cao:2017ndf,Chiaramello:2020ehz,Albanesi:2023bgi,Albanesi:2022xge,Riemenschneider:2021ppj,Chiaramello:2020ehz,Ramos-Buades:2021adz,Liu:2023ldr,Huerta:2016rwp,Huerta:2017kez,Joshi:2022ocr,Setyawati:2021gom,Wang:2023ueg,Islam:2021mha,Carullo:2023kvj,Nagar:2021gss,Tanay:2016zog,Paul:2024ujx,Manna:2024ycx,Gamboa:2024hli,Gamboa:2024imd,Buonanno:1998gg,Buonanno:2000ef}, and estimate eccentricity in LIGO-Virgo-KAGRA BBH merger events~\cite{Romero-Shaw:2020thy,Gayathri:2020coq,Gamba:2021gap,Ramos-Buades:2023yhy,Gupte:2024jfe}. Parallel to these advancements, efforts are underway to standardize the definition of eccentricity in gravitational waveforms~\cite{blanchet:2013haa,Mroue:2010re, Healy:2017zqj,Mora:2002gf,Ramos-Buades:2021adz,Islam:2021mha,Ramos-Buades:2019uvh,Shaikh:2023ypz,Knee:2022hth,Boschini:2024scu,Islam:2025oiv} and identify simple relationships that connect waveform features to eccentricity. Despite this progress, developing accurate and efficient waveform and remnant models for eccentric BBH systems remains a key challenge in the community.

Early studies using Newtonian~\cite{Yunes:2009yz} and PN frameworks~\cite{VanDenBroeck:2006qu,Arun:2007qv,Seto:2001pg,2012ApJ74537V} have hinted at a simple decomposition of the spherical harmonic modes of GW signals from eccentric binaries. These studies suggest that each spherical harmonic mode can be expressed as a sum of monotonic eccentric harmonics, whose frequencies and phases follow a simple hierarchical structure. Recently, Ref.~\cite{Patterson:2024vbo,Islam2025InPrep} introduced efficient frameworks for extracting these harmonics directly from waveform data. The phenomenology of these harmonics is further explored in the companion paper. However, properly extracting the harmonics for each point in the parameter space is computationally expensive (as detailed below). In this work, we therefore develop a fast and data-driven (surrogate) model for the eccentric harmonics extracted in Ref.~\cite{Islam2025InPrep}.

\begin{figure*}
\includegraphics[width=\textwidth]{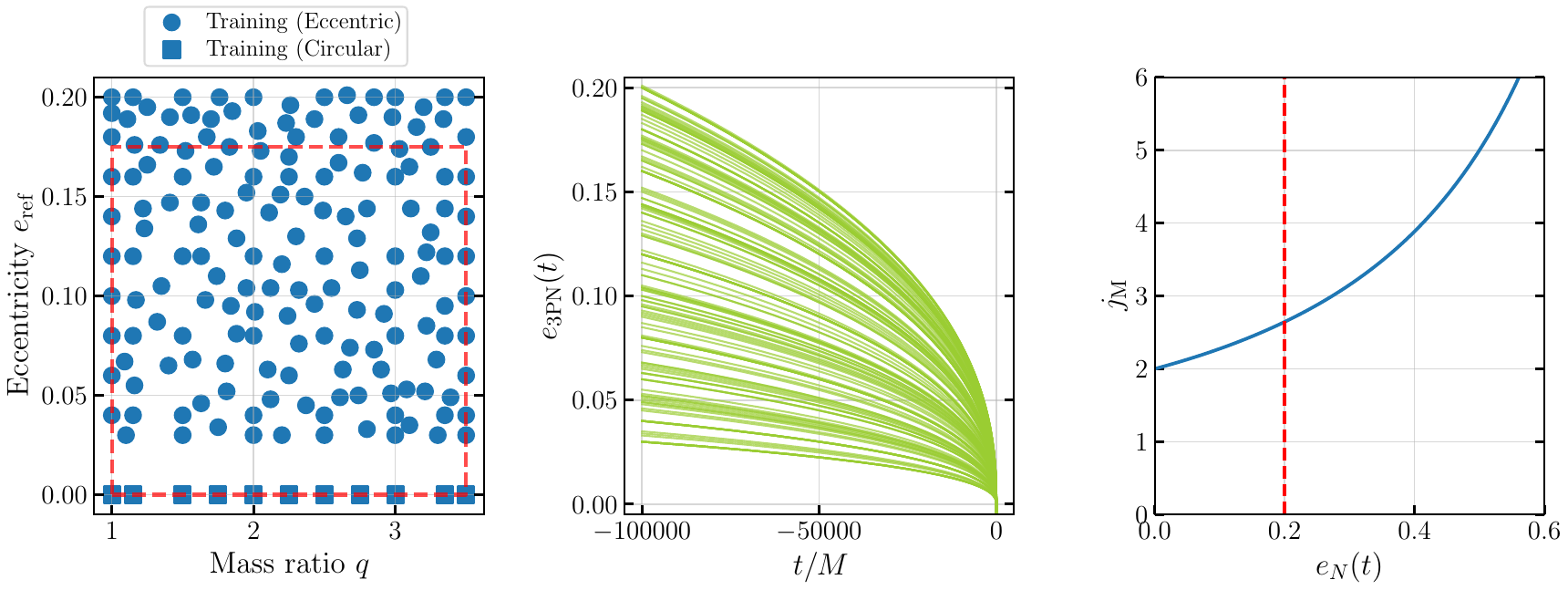}
\caption{(\textit{Left panel}:) We show the training points (both for the circular and eccentric cases) used in building the \texttt{gwharmone} model with \texttt{TEOBResumS} eccentric waveforms. The red dashed rectangle indicates our target parameter space. (\textit{Middle panel}): We show the $3$PN eccentricity evolution of these systems. (\textit{Right panel}): We show the Newtonian expectation of the leading eccentric harmonic $j_{\rm M}$ as a function of the Newtonian eccentricity. Red dashed vertical line shows the boundary of our training region.}
\label{fig:training}
\end{figure*}

\noindent {\textbf{\textit{Full order model}}.}
The structure of these eccentric harmonics $\{h_{\ell m, j}\}$ within any spherical harmonic $h_{\ell,m}(t; \boldsymbol\lambda)$ is the following:
\begin{align}
h_{\ell,m}(t; \boldsymbol\lambda) &= \sum_{j=1}^\infty h_{\ell m, j}(t; \boldsymbol\lambda).
\label{hmodes}
\end{align}
Here, $\boldsymbol{\lambda}$ is the set of parameters that describe the binary, given by:  $\boldsymbol{\lambda} = \{q, \boldsymbol{\chi}_1, \boldsymbol{\chi}_2, e_{\rm ref}, l_{\rm ref} \}$,  
where $q$ is the mass ratio, $\boldsymbol{\chi}_1$ and $\boldsymbol{\chi}_2$ are the three-dimensional spin vectors of the two black holes, and $e_{\rm ref}$ and $l_{\rm ref}$ denote the eccentricity and mean anomaly, respectively, both defined at a chosen reference time or frequency. Because we work in the non-spinning limit, $\boldsymbol{\chi}_1 = \boldsymbol{\chi}_2 = [0,0,0]$ for us. We simulate the first four eccentric harmonics indexed by $j = [1,2,3,4]$, from the dominant quadrupolar mode (i.e. $\ell=2$, $m=2$) using the \texttt{TEOBResumS} approximant~\cite{Chiaramello:2020ehz} via the \texttt{gwMiner} package~\cite{Islam2025InPrep}~\footnote{\href{https://github.com/tousifislam/gwMiner}{https://github.com/tousifislam/gwMiner} (available on request)} (\texttt{full order model}), where $i$ indexes the training waveforms. \texttt{TEOBResumS}~\footnote{\href{https://bitbucket.org/teobresums/teobresums}{https://bitbucket.org/teobresums/teobresums/} package (available at the \texttt{attic/eccentric} branch)} is based on the effective one-body (EOB) formalism and is tuned to NR data, while \texttt{gwMiner} is a package that provides an efficient computational framework based on signal processing and SVD to extract eccentric harmonics from a given eccentric approximant (in our case, from \texttt{TEOBResumS}~\cite{Chiaramello:2020ehz}). Typically, the \texttt{full order model} requires $\sim 50$ \texttt{TEOBResumS} waveform generations for the same $[q,e_{\rm ref}]$ but with different values of $l_{\rm ref}$ to calculate the eccentric harmonics.

\noindent {\textbf{\textit{Snapshot data}}.}
Our target parameter space is $q \in [1,3.5]$ and $e_{\rm ref} \in [0.03,0.175]$. We generate eccentric harmonics at a total of 162 points within the range $q \in [1,3.5]$ and $e_{\rm ref} \in [0.03,0.2]$ (at the start of the waveform). While generating the harmonics, we fix mean anomaly $l_{\rm ref}=\pi$ (as the same choice is made within \texttt{TEOBResumS}). We then use the phenomenologically observed universal relation between eccentric harmonics at different $l_{\rm ref}$ values (for the same $q$ and $e_{\rm ref}$) to model the mean anomaly direction. Details of these relations are provided in Ref.~\cite{Islam2025InPrep}.

Note that while generating training data, we extend beyond our eccentricity target limit of $e_{\rm ref} = 0.175$. This is because data-driven models typically exhibit larger errors near the boundaries~\cite{varma2019surrogate,Islam:2021mha}, and eccentricity is already a challenging feature to model. Therefore, extending the training range beyond the target range helps to ensure smaller errors throughout the target parameter space. Additionally, we include 11 non-eccentric anchor points generated using the \texttt{TEOBResumS} approximant (in its circular limit). 
All our waveforms have a duration of $100,000M$~\footnote{This implies that the minimum detector-frame mass for which we can generate waveforms starting from $20$ Hz is $\sim 12M_{\odot}$ for $q=1$ and $\sim 19.5M_{\odot}$ for $q=4$.}. To ensure consistency in the training data, we cast the raw eccentric harmonics $\{h_{j,i}(t_i)\}$ onto a common time grid $t \in [t_{\rm min},t_{\rm max}]$, where $t_{\rm min} = \max (\{\min(t_i)\})$ and $t_{\rm max} = \min (\{\max(t_i)\})$.

We use a uniform time step of $0.1M$. Additionally, we align the waveforms such that the initial phase of each harmonic (and thereby the full waveform) is set to zero at the beginning of the waveform.
In Figure~\ref{fig:training}, we show the training points (both circular and eccentric) along with our target parameter space, highlighted by a red dashed rectangle. Furthermore, we depict the eccentricity evolution of all binaries used in training, estimated using the 3PN approximated analytical expression obtained from Ref.~\cite{Moore:2016qxz}. We also compute the Newtonian expectation of the leading eccentric harmonic $j_{\rm M}=2(1+e_{\rm N})^{1.1954}/(1-e_{\rm N}^2)^{1.5}$ (that has most of the radiated power) as a function of the Newtonian eccentricity $e_{\rm N}$ following expressions provided in Ref.~\cite{Wen:2002km}. For our case, we find that $j=2$ would always contain most of the radiation.

\noindent {\textbf{\textit{Hierarchical modeling}}.}
Our goal is to model the extracted eccentric harmonics as a function of $q$, $e_{\rm ref}$ and $l_{\rm ref}$ while ensuring that any small noisy features in the extracted harmonics are excluded, resulting in smooth model outputs. Since the circular waveform can be generated directly from \texttt{TEOBResumS} (without requiring the \texttt{gwMiner} package) and is not affected by the small-scale noise that often appears in the extracted eccentric harmonics, we adopt a hierarchical modeling framework. We start with the circular base model and gradually incorporate corrections to account for eccentricity effects. Our modeling framework involves constructing sub-models for several key pieces of the waveform. Below, we outline the steps.

First, we decompose the circular waveform into its amplitude, $A_{22}^{\rm cir}(t;q)$, and phase, $\phi_{22}^{\rm cir}(t;q)$. We then construct models for these quantities, denoted as $A_{22}^{\rm cir,M}(t;q)$ and $\phi_{22}^{\rm cir,M}(t;q)$, respectively. Throughout the paper, we will use superscript `$\rm M$' to denote a model. Next, for the dominant $j=2$ harmonic, we compute the amplitude correction relative to the circular expectation:
\begin{equation}
\Delta A_{22,2}(t) = A_{22,2}(t) - A_{22}^{\rm cir,M}(t),
\end{equation}
and build a model for it. Note that we use the modeled circular amplitude to subtract the circular expectation. For all other harmonics ($j \neq 2$), we directly model the full amplitudes, $A_{22,j}(t)$.

For the $j=2$ harmonic, we compute the additional phase relative to the circular expectation (again using the modeled circular phase):
\begin{equation}
\Delta \phi_{22,2}(t) = \phi_{22,2}(t) - \phi_{22}^{\rm cir,M}(t).
\end{equation}
We then construct a model for this additional phase as a function of $q$ and $e_{\rm ref}$, denoted as $\Delta \phi_{22,2}^{\rm M}(t; q, e_{\rm ref})$.
For the other harmonics ($j \neq 2$), we define the extra phases as:
\begin{equation}
\Delta \phi_{22,j}(t) = \phi_{22,j}(t) - \frac{j\phi_{22}^{\rm cir,M}(t)}{2} - \Delta \phi_{22,2}^{\rm M}(t),
\end{equation}
and subsequently build models for these extra phases, denoted as $\Delta \phi_{22,j}^{\rm M}(t)$. We call the resultant final waveform model \texttt{gwharmone}.

\noindent {\textbf{\textit{Model order reduction}}.}
For each data piece $\{x_i\}$, we construct a matrix $X$ with dimensions $n \times m$, where $n$ is the number of training data points and $m$ is the number of points in the time grid. We then apply SVD, using the \texttt{scipy.linalg.svd} module~\footnote{\href{https://docs.scipy.org/doc/scipy/reference/generated/scipy.linalg.svd.html}{https://docs.scipy.org/doc/scipy/reference/generated/scipy.linalg.svd.html}}, to factorize $X$ into three matrices: $U$, $\Sigma$, and $V^T$, with dimensions $m \times m$, $n \times n$, and $m \times n$, respectively: $X = U \Sigma V^T$~\cite{10.5555/1538674}. Similar model reduction techniques have recently been used in waveform modelling and detection (e.g. see ~\cite{varma2019surrogate,Islam:2021mha,Pathak:2024zgo,Wadekar:2024zdq,Roulet:2019hzy}).
Here, $V^T$ provides the `modes' or SVD basis vectors $\{v_1, v_2, \dots, v_n\}$ while (diagonal) matrix $\Sigma$ contains non-negative real numbers $\{\lambda_1, \lambda_2, \dots, \lambda_n\}$ (known as \textit{singular values}) that indicates the significance of the corresponding SVD vectors. Each data piece $x_i$ can now be approximated as a linear superposition of the SVD basis vectors:
\begin{equation}
x_i = \sum_{i^{\prime}=1}^{n} a_{i^{\prime}}(q, e_{\rm ref}) v_{i^{\prime}},
\end{equation}
where $a_{i^{\prime}}(q, e_{\rm ref})$ are the coefficients to the basis vectors.
For each piece of data, instead of utilizing several basis vectors, we restrict ourselves to using only the most important two or three (and in some cases, even just one) basis vector - thereby reducing the model order. This ensures that we primarily model the most significant features of the data while minimizing the influence of noise, which is typically captured by the later basis vectors.

\begin{figure}
\includegraphics[width=\columnwidth]{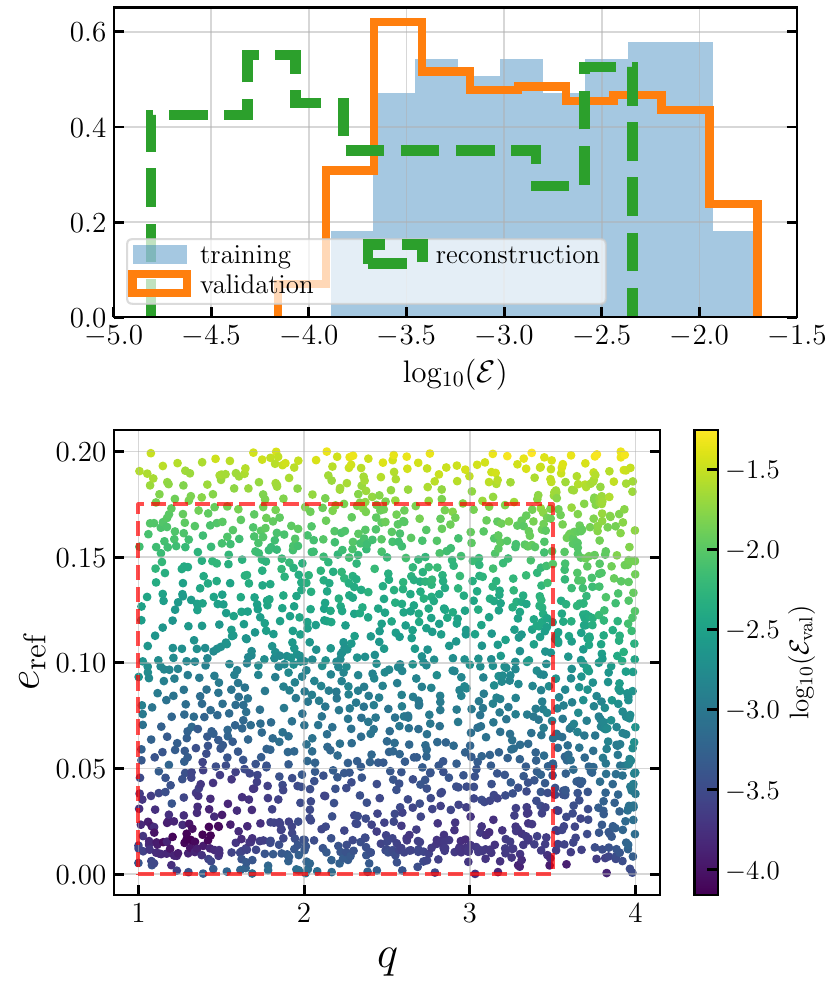}
\caption{\textit{Upper panel:} We show the time-domain errors of the \texttt{gwharmone} model during the training and validation stages for eccentricities less than $0.175$ (our intended parameter space), compared against the baseline \texttt{TEOBResumS} eccentric waveforms. For comparison, we also present the reconstruction error of the eccentric harmonics obtained using the SVD method. \textit{Bottom panel:} Furthermore, we show the error as a function of the parameter space for all validation points, including those within the target parameter space (shown as a red rectangle).}
\label{fig:time_domain_errors}
\end{figure}

\noindent {\textbf{\textit{Denoising certain SVD bases}}.}
To ensure smoothness, we further inspect each SVD basis visually (despite the cumbersome nature of this task) and remove any noisy artifacts using smoothing splines. While these choices slightly reduce model accuracy, we prioritize smoothness over retaining noisy features in the final model. 
Additionally, we find that the basis vectors obtained for the data components $\Delta \phi_{22,1}(t)$ and $\Delta \phi_{22,3}(t)$ exhibit significant noise, making it challenging to smooth them effectively. As a result, we choose not to model these data pieces using the SVD basis. Instead, we employ phenomenologically derived phase relations between eccentric harmonics to reconstruct these two phases at a later stage (using the framework given in Ref.~\cite{Islam2025InPrep}).

\noindent {\textbf{\textit{Regression and Interpolation}}.}
Once we determine the coefficients as functions of the parameter space, we construct data-driven models using two regression methods: Gaussian process regression (GPR) and random forest (RF), to predict the SVD basis coefficients for each data piece. Additionally, we employ radial basis function (RBF) interpolators for mesh-free interpolation. 
For each scenario, we test multiple kernel and hyperparameter choices while simultaneously verifying that the resulting regression/interpolation models extrapolate well beyond the training range. We find that in all cases, GPR outperforms the other methods. Consequently, we use GPR to fit the SVD basis coefficients for all data pieces, except for the circular case. For the circular waveform (both amplitude and phase), we instead use cubic splines.
Our implementation of GPR, RF, RBF and cubic splines are based on \texttt{scikit-learn} modules~\footnote{\href{https://scikit-learn.org/}{https://scikit-learn.org/}}.

\noindent {\textbf{\textit{Model evaluation}}.}
The model takes a specific mass ratio $q = q_0$, reference eccentricity $e_{\rm ref} = e_{\rm ref,0}$ and reference mean anomaly $l_{\rm ref} = l_{\rm ref,0}$ as input. It then utilizes splines (for the circular case) and GPRs (for the eccentric corrections) to predict all necessary SVD coefficients (for $l_{\rm ref}=\pi$). We then utilize the phenomenologically derived universal relation between SVD coefficients and mean anomaly values to obtain the corresponding coefficients for the input $l_{\rm ref}$ (using the framework given in Ref.~\cite{Islam2025InPrep}).
These coefficients are then multiplied by their corresponding basis vectors to compute the following data pieces: (i) The circular waveform components: $A_{22}^{\rm cir}(t;q)$ and $\phi_{22}^{\rm cir}(t;q)$; (ii) The eccentric corrections for the $j=2$ harmonic: $\Delta A_{22,2}(t)$ and $\Delta \phi_{22,2}(t)$; and (iii) The eccentric harmonic amplitudes: $\Delta \phi_{22,4}(t)$, $A_{22,1}(t;q)$, $A_{22,3}(t;q)$, and $A_{22,4}(t;q)$.   
For the $j=2$ harmonic, we obtain the total amplitude as: $A_{22,2}(t) = A_{22}^{\rm cir}(t) + \Delta A_{22,2}(t)$.
The phases of the $j=2$ and $j=4$ harmonics are computed as:
\begin{align}
\phi_{22,2}(t) &= \phi_{22}^{\rm cir}(t) + \Delta \phi_{22,2}(t),\\
\phi_{22,4}(t) &= 2\phi_{22}^{\rm cir}(t) + \Delta \phi_{22,2}(t) + \Delta \phi_{22,4}(t).
\end{align}
Next, we compute the secular orbital phase and the eccentric phase as:
\begin{align}
\phi_{\lambda} &= \frac{\phi_{22,4}(t) - \phi_{22,2}(t)}{2},\\
\phi_{\rm ecc} &= \phi_{22,2}(t) - 2\phi_{\lambda}(t).
\end{align}
At this stage, we use the phase relations of the eccentric harmonics (provided in Eq.~(24) of Ref.~\cite{Islam2025InPrep}) to determine the phases for the $j=1$ and $j=3$ harmonics: 
\begin{align}
\phi_{22,1}(t) &= \phi_{\lambda} + \phi_{\rm ecc},\\
\phi_{22,3}(t) &= 3\phi_{\lambda} + \phi_{\rm ecc} + \pi.
\end{align}
Finally, once we obtain the amplitudes and phases of each harmonic, we combine them to construct the full complex-valued waveform modes.

\begin{figure}
\includegraphics[width=\columnwidth]{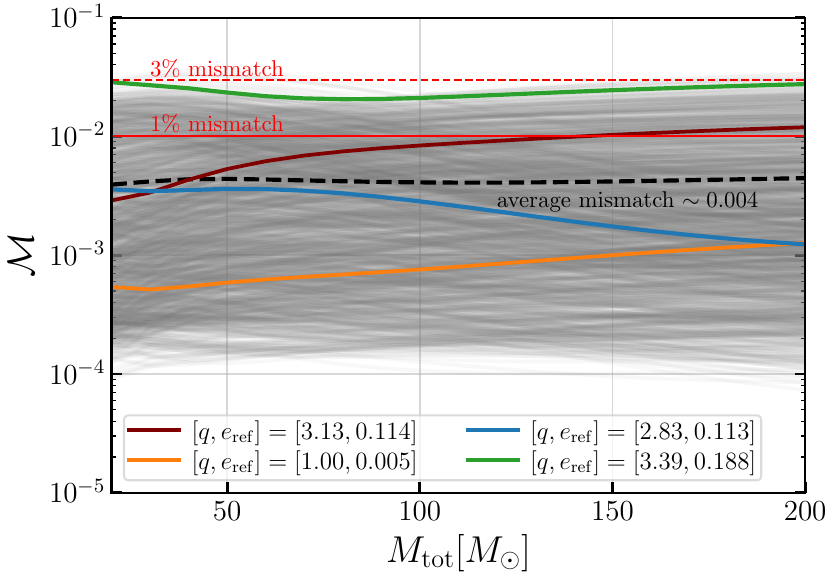}
\caption{We show the frequency-domain mismatches of the \texttt{gwharmone} model when compared against the baseline \texttt{TEOBResumS} eccentric waveforms for eccentricities less than $0.175$, assuming the advanced LIGO noise sensitivity curve. The solid black line represents the average mismatch of the model, while the red lines indicate the $1\%$ and $3\%$ mismatch thresholds.
}
\label{fig:mismatches}
\end{figure}

\noindent {\textbf{\textit{Model accuracy}}.}
To investigate the model accuracy, we calculate both time-domain and frequency-domain errors of the \texttt{gwharmone} predictions ($h_{\rm M}$) with respect to the base \texttt{TEOBResumS} model ($h_{\rm EOB}$). Time-domain errors are obtained by computing a relative $L_2$-norm error (denoted by $\mathcal{E}$) between the two waveforms while performing a time and phase optimization (as defined in Ref.~\cite{Blackman:2017dfb}).
This is similar to calculating a mismatch over white noise but entirely in the time domain. We compute these errors in three cases: (i) error between \texttt{gwharmone} predictions and \texttt{TEOBResumS} waveforms at the 162 training points (\textit{training errors}); (ii) error between \texttt{gwharmone} predictions and \texttt{TEOBResumS} waveforms at 1751 validation points randomly distributed throughout the parameter space (\textit{validation errors}); and (iii) error between the total waveform obtained by summing the first four eccentric harmonics in the \texttt{gwMiner} output and \texttt{TEOBResumS} waveforms at all training points (\textit{reconstruction errors}). The latter also serves as a baseline numerical error for evaluating the model's accuracy.

\begin{figure}
\includegraphics[width=\columnwidth]{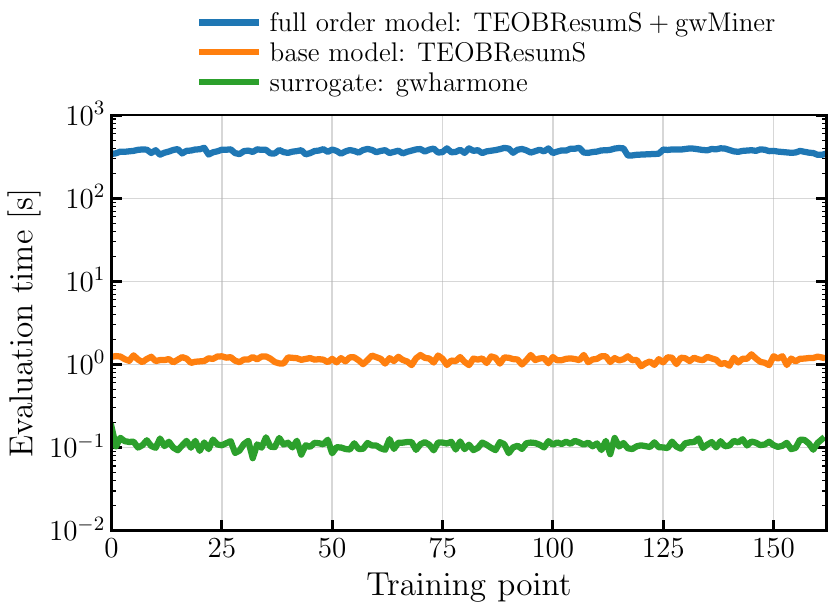}
\caption{We present the evaluation times for the full-order model (\texttt{TEOBResumS} with \texttt{gwMiner}), the base model (\texttt{TEOBResumS}) which only provides the full eccentric waveform (without the harmonics), and our surrogate model (\texttt{gwharmone}) which predicts both the eccentric harmonics and the quadrupolar mode, at the parameter points used for training.}
\label{fig:harmone_timings}
\end{figure}

In Figure~\ref{fig:time_domain_errors}, the upper panel shows a histogram of these three errors for waveforms with initial eccentricities below $0.175$ (i.e. within our target space). We find that the validation and training errors are comparable, with validation errors having a slightly lower average than training errors. However, both training and validation errors are larger than the reconstruction errors. Overall, our validation errors (which are the true indicators of model accuracy) remain around $10^{-3}$ for most of the parameter space, suggesting that the model is quite accurate. In the lower panel of the same figure, we show the time-domain errors as a function of the parameter space, including regions where the model extrapolates beyond the training range. We find that the model performs well within the target parameter space and extrapolates accurately in the $q$ direction for moderate eccentricities, maintaining accuracy up to $q=4$.

Next, we compute the frequency-domain mismatches (as defined in Ref.~\cite{Cutler:1994ys}) between \texttt{gwharmone} predictions and \texttt{TEOBResumS} waveforms for all 1570 validation points for which initial eccentricity is less than $0.175$, assuming the Advanced LIGO sensitivity noise curve~\cite{KAGRA:2013rdx} for total masses varying from $M=20M_{\odot}$ to $M=200M_{\odot}$~\footnote{For LISA or other 3G detectors, waveform models capable of generating longer harmonics are needed.}. The results are shown in Figure~\ref{fig:mismatches}. 
We find that in most cases, the mismatches remain below $10^{-2}$, often reaching values as low as $10^{-4}$. The average mismatch is approximately $0.004$. However, for high eccentricity cases ($e_{\rm ref} \geq 0.15$), mismatches increase beyond $10^{-2}$ but do not exceed the $3\%$ mismatch threshold. This suggests that while our model achieves high accuracy across most of the intended parameter space, there is room for improvement at higher eccentricities. We leave further refinement for future work.

\noindent {\textbf{\textit{Model evaluation cost}}.}
We compute the evaluation times for the full-order model (\texttt{TEOBResumS} with \texttt{gwMiner}), the base model (\texttt{TEOBResumS}), and our surrogate model (\texttt{gwharmone}), which predicts both the eccentric harmonics and the quadrupolar mode, at the parameter points used for training (see Figure~\ref{fig:harmone_timings}). We find that the surrogate model (\texttt{gwharmone}) takes, on average, $0.1$ seconds to evaluate, while the base model takes approximately $1$ second (providing only the full quadrupolar mode). In contrast, the full-order model (which provides both the quadrupolar mode and the constituent eccentric harmonics) takes $\sim 300$ seconds to evaluate. Thus, our surrogate model is an order of magnitude faster than the base \texttt{TEOBResumS} model and at least three orders of magnitude faster than the full-order model. This demonstrates that \texttt{gwharmone} not only offers an accurate and efficient way of generating eccentric harmonics but also provides an efficient surrogate for the full quadrupolar mode of the \texttt{TEOBResumS} model.

\noindent {\textbf{\textit{Summary and Future Work}}.}
To summarize, we present the first data-driven, faithful, and fast surrogate model, \texttt{gwharmone}, for the dominant four eccentric harmonics, $j=[1,2,3,4]$, in non-spinning eccentric BBH merger quadrupolar waveforms. These harmonics are smooth, monotonic functions of time and frequency, unlike the complete quadrupole mode, which exhibits oscillatory behavior. The model is trained for mass ratios $q \in [1,3.5]$ and eccentricities $e_{\rm ref} \in [0,0.2]$ (at the start of the waveform). Our average frequency-domain model mismatches are approximately $0.004$, while the evaluation cost is one to three orders of magnitude lower than that of the base and full-order models, making it a suitable choice for routine detection and data analysis. This model also opens up the possibility of building efficient search algorithms based on the eccentric harmonics similar to those presented in Refs.~\cite{Wadekar:2024zdq} and developing faster eccentric data analysis frameworks akin to Refs.~\cite{Zackay:2018qdy, Roulet:2024hwz} by utilizing the monotonicity of the eccentric harmonics. These advancements will significantly aid in the detection and interpretation of eccentric binaries using GWs, and we will pursue these directions in the near future. While our model currently focuses on non-spinning eccentric systems, the harmonic extraction and modeling framework can also be applied to eccentric, non-precessing systems. One of our immediate goals is therefore to incorporate spins into the \texttt{gwharmone} model. Finally, in the future, we also aim to build similar models based on other state-of-the-art waveform frameworks, such as the \texttt{SEOBNR} models, as well as NR data.

\noindent {\textbf{\textit{Model availability}}.}
The model will be publicly available through the \texttt{gwModels} package, hosted at \href{https://github.com/tousifislam/gwModels}{https://github.com/tousifislam/gwModels}. It can also be accessed through the \texttt{PyPI} package manager.

\begin{acknowledgments}
We thank Steve Fairhurst, Ben Patterson, Scott Field, Peter James Nee and Antoni Ramos-Buades for useful discussions and Aldo Gamboa for helpful comments on an earlier version of the draft.
This research was supported in part by the National Science Foundation under Grant No. NSF PHY-2309135 and the Simons Foundation (216179, LB). 
Use was made of computational facilities purchased with funds from the National Science Foundation (CNS-1725797) and administered by the Center for Scientific Computing (CSC). The CSC is supported by the California NanoSystems Institute and the Materials Research Science and Engineering Center (MRSEC; NSF DMR 2308708) at UC Santa Barbara.
JR acknowledges support from the Sherman Fairchild Foundation.
\end{acknowledgments}

\bibliography{References}

\begin{thebibliography}{86}%
\makeatletter
\providecommand \@ifxundefined [1]{%
 \@ifx{#1\undefined}
}%
\providecommand \@ifnum [1]{%
 \ifnum #1\expandafter \@firstoftwo
 \else \expandafter \@secondoftwo
 \fi
}%
\providecommand \@ifx [1]{%
 \ifx #1\expandafter \@firstoftwo
 \else \expandafter \@secondoftwo
 \fi
}%
\providecommand \natexlab [1]{#1}%
\providecommand \enquote  [1]{``#1''}%
\providecommand \bibnamefont  [1]{#1}%
\providecommand \bibfnamefont [1]{#1}%
\providecommand \citenamefont [1]{#1}%
\providecommand \href@noop [0]{\@secondoftwo}%
\providecommand \href [0]{\begingroup \@sanitize@url \@href}%
\providecommand \@href[1]{\@@startlink{#1}\@@href}%
\providecommand \@@href[1]{\endgroup#1\@@endlink}%
\providecommand \@sanitize@url [0]{\catcode `\\12\catcode `\$12\catcode `\&12\catcode `\#12\catcode `\^12\catcode `\_12\catcode `\%12\relax}%
\providecommand \@@startlink[1]{}%
\providecommand \@@endlink[0]{}%
\providecommand \url  [0]{\begingroup\@sanitize@url \@url }%
\providecommand \@url [1]{\endgroup\@href {#1}{\urlprefix }}%
\providecommand \urlprefix  [0]{URL }%
\providecommand \Eprint [0]{\href }%
\providecommand \doibase [0]{http://dx.doi.org/}%
\providecommand \selectlanguage [0]{\@gobble}%
\providecommand \bibinfo  [0]{\@secondoftwo}%
\providecommand \bibfield  [0]{\@secondoftwo}%
\providecommand \translation [1]{[#1]}%
\providecommand \BibitemOpen [0]{}%
\providecommand \bibitemStop [0]{}%
\providecommand \bibitemNoStop [0]{.\EOS\space}%
\providecommand \EOS [0]{\spacefactor3000\relax}%
\providecommand \BibitemShut  [1]{\csname bibitem#1\endcsname}%
\let\auto@bib@innerbib\@empty
\bibitem [{\citenamefont {Harry}(2010)}]{Harry:2010zz}%
  \BibitemOpen
  \bibfield  {author} {\bibinfo {author} {\bibfnamefont {Gregory~M.}\ \bibnamefont {Harry}} (\bibinfo {collaboration} {LIGO Scientific}),\ }\bibfield  {title} {\enquote {\bibinfo {title} {{Advanced LIGO: The next generation of gravitational wave detectors}},}\ }\href {\doibase 10.1088/0264-9381/27/8/084006} {\bibfield  {journal} {\bibinfo  {journal} {Class. Quant. Grav.}\ }\textbf {\bibinfo {volume} {27}},\ \bibinfo {pages} {084006} (\bibinfo {year} {2010})}\BibitemShut {NoStop}%
\bibitem [{\citenamefont {Acernese}\ \emph {et~al.}(2015)\citenamefont {Acernese} \emph {et~al.}}]{VIRGO:2014yos}%
  \BibitemOpen
  \bibfield  {author} {\bibinfo {author} {\bibfnamefont {F.}~\bibnamefont {Acernese}} \emph {et~al.} (\bibinfo {collaboration} {VIRGO}),\ }\bibfield  {title} {\enquote {\bibinfo {title} {{Advanced Virgo: a second-generation interferometric gravitational wave detector}},}\ }\href {\doibase 10.1088/0264-9381/32/2/024001} {\bibfield  {journal} {\bibinfo  {journal} {Class. Quant. Grav.}\ }\textbf {\bibinfo {volume} {32}},\ \bibinfo {pages} {024001} (\bibinfo {year} {2015})},\ \Eprint {http://arxiv.org/abs/1408.3978} {arXiv:1408.3978 [gr-qc]} \BibitemShut {NoStop}%
\bibitem [{\citenamefont {Akutsu}\ \emph {et~al.}(2021)\citenamefont {Akutsu} \emph {et~al.}}]{KAGRA:2020tym}%
  \BibitemOpen
  \bibfield  {author} {\bibinfo {author} {\bibfnamefont {T.}~\bibnamefont {Akutsu}} \emph {et~al.} (\bibinfo {collaboration} {KAGRA}),\ }\bibfield  {title} {\enquote {\bibinfo {title} {{Overview of KAGRA: Detector design and construction history}},}\ }\href {\doibase 10.1093/ptep/ptaa125} {\bibfield  {journal} {\bibinfo  {journal} {PTEP}\ }\textbf {\bibinfo {volume} {2021}},\ \bibinfo {pages} {05A101} (\bibinfo {year} {2021})},\ \Eprint {http://arxiv.org/abs/2005.05574} {arXiv:2005.05574 [physics.ins-det]} \BibitemShut {NoStop}%
\bibitem [{\citenamefont {Abbott}\ \emph {et~al.}(2019)\citenamefont {Abbott} \emph {et~al.}}]{LIGOScientific:2018mvr}%
  \BibitemOpen
  \bibfield  {author} {\bibinfo {author} {\bibfnamefont {B.~P.}\ \bibnamefont {Abbott}} \emph {et~al.} (\bibinfo {collaboration} {LIGO Scientific, Virgo}),\ }\bibfield  {title} {\enquote {\bibinfo {title} {{GWTC-1: A Gravitational-Wave Transient Catalog of Compact Binary Mergers Observed by LIGO and Virgo during the First and Second Observing Runs}},}\ }\href {\doibase 10.1103/PhysRevX.9.031040} {\bibfield  {journal} {\bibinfo  {journal} {Phys. Rev. X}\ }\textbf {\bibinfo {volume} {9}},\ \bibinfo {pages} {031040} (\bibinfo {year} {2019})},\ \Eprint {http://arxiv.org/abs/1811.12907} {arXiv:1811.12907 [astro-ph.HE]} \BibitemShut {NoStop}%
\bibitem [{\citenamefont {Abbott}\ \emph {et~al.}(2021{\natexlab{a}})\citenamefont {Abbott} \emph {et~al.}}]{LIGOScientific:2020ibl}%
  \BibitemOpen
  \bibfield  {author} {\bibinfo {author} {\bibfnamefont {R.}~\bibnamefont {Abbott}} \emph {et~al.} (\bibinfo {collaboration} {LIGO Scientific, Virgo}),\ }\bibfield  {title} {\enquote {\bibinfo {title} {{GWTC-2: Compact Binary Coalescences Observed by LIGO and Virgo During the First Half of the Third Observing Run}},}\ }\href {\doibase 10.1103/PhysRevX.11.021053} {\bibfield  {journal} {\bibinfo  {journal} {Phys. Rev. X}\ }\textbf {\bibinfo {volume} {11}},\ \bibinfo {pages} {021053} (\bibinfo {year} {2021}{\natexlab{a}})},\ \Eprint {http://arxiv.org/abs/2010.14527} {arXiv:2010.14527 [gr-qc]} \BibitemShut {NoStop}%
\bibitem [{\citenamefont {Abbott}\ \emph {et~al.}(2021{\natexlab{b}})\citenamefont {Abbott} \emph {et~al.}}]{LIGOScientific:2021usb}%
  \BibitemOpen
  \bibfield  {author} {\bibinfo {author} {\bibfnamefont {R.}~\bibnamefont {Abbott}} \emph {et~al.} (\bibinfo {collaboration} {LIGO Scientific, VIRGO}),\ }\bibfield  {title} {\enquote {\bibinfo {title} {{GWTC-2.1: Deep Extended Catalog of Compact Binary Coalescences Observed by LIGO and Virgo During the First Half of the Third Observing Run}},}\ }\href@noop {} {\  (\bibinfo {year} {2021}{\natexlab{b}})},\ \Eprint {http://arxiv.org/abs/2108.01045} {arXiv:2108.01045 [gr-qc]} \BibitemShut {NoStop}%
\bibitem [{\citenamefont {Abbott}\ \emph {et~al.}(2021{\natexlab{c}})\citenamefont {Abbott} \emph {et~al.}}]{LIGOScientific:2021djp}%
  \BibitemOpen
  \bibfield  {author} {\bibinfo {author} {\bibfnamefont {R.}~\bibnamefont {Abbott}} \emph {et~al.} (\bibinfo {collaboration} {LIGO Scientific, VIRGO, KAGRA}),\ }\bibfield  {title} {\enquote {\bibinfo {title} {{GWTC-3: Compact Binary Coalescences Observed by LIGO and Virgo During the Second Part of the Third Observing Run}},}\ }\href@noop {} {\  (\bibinfo {year} {2021}{\natexlab{c}})},\ \Eprint {http://arxiv.org/abs/2111.03606} {arXiv:2111.03606 [gr-qc]} \BibitemShut {NoStop}%
\bibitem [{\citenamefont {Rodriguez}\ \emph {et~al.}(2018{\natexlab{a}})\citenamefont {Rodriguez}, \citenamefont {Amaro-Seoane}, \citenamefont {Chatterjee},\ and\ \citenamefont {Rasio}}]{Rodriguez:2017pec}%
  \BibitemOpen
  \bibfield  {author} {\bibinfo {author} {\bibfnamefont {Carl~L.}\ \bibnamefont {Rodriguez}}, \bibinfo {author} {\bibfnamefont {Pau}\ \bibnamefont {Amaro-Seoane}}, \bibinfo {author} {\bibfnamefont {Sourav}\ \bibnamefont {Chatterjee}}, \ and\ \bibinfo {author} {\bibfnamefont {Frederic~A.}\ \bibnamefont {Rasio}},\ }\bibfield  {title} {\enquote {\bibinfo {title} {{Post-Newtonian Dynamics in Dense Star Clusters: Highly-Eccentric, Highly-Spinning, and Repeated Binary Black Hole Mergers}},}\ }\href {\doibase 10.1103/PhysRevLett.120.151101} {\bibfield  {journal} {\bibinfo  {journal} {Phys. Rev. Lett.}\ }\textbf {\bibinfo {volume} {120}},\ \bibinfo {pages} {151101} (\bibinfo {year} {2018}{\natexlab{a}})},\ \Eprint {http://arxiv.org/abs/1712.04937} {arXiv:1712.04937 [astro-ph.HE]} \BibitemShut {NoStop}%
\bibitem [{\citenamefont {Rodriguez}\ \emph {et~al.}(2018{\natexlab{b}})\citenamefont {Rodriguez}, \citenamefont {Amaro-Seoane}, \citenamefont {Chatterjee}, \citenamefont {Kremer}, \citenamefont {Rasio}, \citenamefont {Samsing}, \citenamefont {Ye},\ and\ \citenamefont {Zevin}}]{Rodriguez:2018pss}%
  \BibitemOpen
  \bibfield  {author} {\bibinfo {author} {\bibfnamefont {Carl~L.}\ \bibnamefont {Rodriguez}}, \bibinfo {author} {\bibfnamefont {Pau}\ \bibnamefont {Amaro-Seoane}}, \bibinfo {author} {\bibfnamefont {Sourav}\ \bibnamefont {Chatterjee}}, \bibinfo {author} {\bibfnamefont {Kyle}\ \bibnamefont {Kremer}}, \bibinfo {author} {\bibfnamefont {Frederic~A.}\ \bibnamefont {Rasio}}, \bibinfo {author} {\bibfnamefont {Johan}\ \bibnamefont {Samsing}}, \bibinfo {author} {\bibfnamefont {Claire~S.}\ \bibnamefont {Ye}}, \ and\ \bibinfo {author} {\bibfnamefont {Michael}\ \bibnamefont {Zevin}},\ }\bibfield  {title} {\enquote {\bibinfo {title} {{Post-Newtonian Dynamics in Dense Star Clusters: Formation, Masses, and Merger Rates of Highly-Eccentric Black Hole Binaries}},}\ }\href {\doibase 10.1103/PhysRevD.98.123005} {\bibfield  {journal} {\bibinfo  {journal} {Phys. Rev. D}\ }\textbf {\bibinfo {volume} {98}},\ \bibinfo {pages} {123005} (\bibinfo {year} {2018}{\natexlab{b}})},\ \Eprint {http://arxiv.org/abs/1811.04926}
  {arXiv:1811.04926 [astro-ph.HE]} \BibitemShut {NoStop}%
\bibitem [{\citenamefont {Samsing}(2018)}]{Samsing:2017xmd}%
  \BibitemOpen
  \bibfield  {author} {\bibinfo {author} {\bibfnamefont {Johan}\ \bibnamefont {Samsing}},\ }\bibfield  {title} {\enquote {\bibinfo {title} {{Eccentric Black Hole Mergers Forming in Globular Clusters}},}\ }\href {\doibase 10.1103/PhysRevD.97.103014} {\bibfield  {journal} {\bibinfo  {journal} {Phys. Rev. D}\ }\textbf {\bibinfo {volume} {97}},\ \bibinfo {pages} {103014} (\bibinfo {year} {2018})},\ \Eprint {http://arxiv.org/abs/1711.07452} {arXiv:1711.07452 [astro-ph.HE]} \BibitemShut {NoStop}%
\bibitem [{\citenamefont {Zevin}\ \emph {et~al.}(2019)\citenamefont {Zevin}, \citenamefont {Samsing}, \citenamefont {Rodriguez}, \citenamefont {Haster},\ and\ \citenamefont {Ramirez-Ruiz}}]{Zevin:2018kzq}%
  \BibitemOpen
  \bibfield  {author} {\bibinfo {author} {\bibfnamefont {Michael}\ \bibnamefont {Zevin}}, \bibinfo {author} {\bibfnamefont {Johan}\ \bibnamefont {Samsing}}, \bibinfo {author} {\bibfnamefont {Carl}\ \bibnamefont {Rodriguez}}, \bibinfo {author} {\bibfnamefont {Carl-Johan}\ \bibnamefont {Haster}}, \ and\ \bibinfo {author} {\bibfnamefont {Enrico}\ \bibnamefont {Ramirez-Ruiz}},\ }\bibfield  {title} {\enquote {\bibinfo {title} {{Eccentric Black Hole Mergers in Dense Star Clusters: The Role of Binary\textendash{}Binary Encounters}},}\ }\href {\doibase 10.3847/1538-4357/aaf6ec} {\bibfield  {journal} {\bibinfo  {journal} {Astrophys. J.}\ }\textbf {\bibinfo {volume} {871}},\ \bibinfo {pages} {91} (\bibinfo {year} {2019})},\ \Eprint {http://arxiv.org/abs/1810.00901} {arXiv:1810.00901 [astro-ph.HE]} \BibitemShut {NoStop}%
\bibitem [{\citenamefont {Zevin}\ \emph {et~al.}(2021)\citenamefont {Zevin}, \citenamefont {Romero-Shaw}, \citenamefont {Kremer}, \citenamefont {Thrane},\ and\ \citenamefont {Lasky}}]{Zevin:2021rtf}%
  \BibitemOpen
  \bibfield  {author} {\bibinfo {author} {\bibfnamefont {Michael}\ \bibnamefont {Zevin}}, \bibinfo {author} {\bibfnamefont {Isobel~M.}\ \bibnamefont {Romero-Shaw}}, \bibinfo {author} {\bibfnamefont {Kyle}\ \bibnamefont {Kremer}}, \bibinfo {author} {\bibfnamefont {Eric}\ \bibnamefont {Thrane}}, \ and\ \bibinfo {author} {\bibfnamefont {Paul~D.}\ \bibnamefont {Lasky}},\ }\bibfield  {title} {\enquote {\bibinfo {title} {{Implications of Eccentric Observations on Binary Black Hole Formation Channels}},}\ }\href {\doibase 10.3847/2041-8213/ac32dc} {\bibfield  {journal} {\bibinfo  {journal} {Astrophys. J. Lett.}\ }\textbf {\bibinfo {volume} {921}},\ \bibinfo {pages} {L43} (\bibinfo {year} {2021})},\ \Eprint {http://arxiv.org/abs/2106.09042} {arXiv:2106.09042 [astro-ph.HE]} \BibitemShut {NoStop}%
\bibitem [{\citenamefont {Samsing}\ \emph {et~al.}(2022)\citenamefont {Samsing}, \citenamefont {Bartos}, \citenamefont {D'Orazio}, \citenamefont {Haiman}, \citenamefont {Kocsis}, \citenamefont {Leigh}, \citenamefont {Liu}, \citenamefont {Pessah},\ and\ \citenamefont {Tagawa}}]{Samsing:2020tda}%
  \BibitemOpen
  \bibfield  {author} {\bibinfo {author} {\bibfnamefont {J.}~\bibnamefont {Samsing}}, \bibinfo {author} {\bibfnamefont {I.}~\bibnamefont {Bartos}}, \bibinfo {author} {\bibfnamefont {D.~J.}\ \bibnamefont {D'Orazio}}, \bibinfo {author} {\bibfnamefont {Z.}~\bibnamefont {Haiman}}, \bibinfo {author} {\bibfnamefont {B.}~\bibnamefont {Kocsis}}, \bibinfo {author} {\bibfnamefont {N.~W.~C.}\ \bibnamefont {Leigh}}, \bibinfo {author} {\bibfnamefont {B.}~\bibnamefont {Liu}}, \bibinfo {author} {\bibfnamefont {M.~E.}\ \bibnamefont {Pessah}}, \ and\ \bibinfo {author} {\bibfnamefont {H.}~\bibnamefont {Tagawa}},\ }\bibfield  {title} {\enquote {\bibinfo {title} {{AGN as potential factories for eccentric black hole mergers}},}\ }\href {\doibase 10.1038/s41586-021-04333-1} {\bibfield  {journal} {\bibinfo  {journal} {Nature}\ }\textbf {\bibinfo {volume} {603}},\ \bibinfo {pages} {237--240} (\bibinfo {year} {2022})},\ \Eprint {http://arxiv.org/abs/2010.09765} {arXiv:2010.09765 [astro-ph.HE]} \BibitemShut {NoStop}%
\bibitem [{\citenamefont {Mroue}\ \emph {et~al.}(2010)\citenamefont {Mroue}, \citenamefont {Pfeiffer}, \citenamefont {Kidder},\ and\ \citenamefont {Teukolsky}}]{Mroue:2010re}%
  \BibitemOpen
  \bibfield  {author} {\bibinfo {author} {\bibfnamefont {Abdul~H.}\ \bibnamefont {Mroue}}, \bibinfo {author} {\bibfnamefont {Harald~P.}\ \bibnamefont {Pfeiffer}}, \bibinfo {author} {\bibfnamefont {Lawrence~E.}\ \bibnamefont {Kidder}}, \ and\ \bibinfo {author} {\bibfnamefont {Saul~A.}\ \bibnamefont {Teukolsky}},\ }\bibfield  {title} {\enquote {\bibinfo {title} {{Measuring orbital eccentricity and periastron advance in quasi-circular black hole simulations}},}\ }\href {\doibase 10.1103/PhysRevD.82.124016} {\bibfield  {journal} {\bibinfo  {journal} {Phys. Rev. D}\ }\textbf {\bibinfo {volume} {82}},\ \bibinfo {pages} {124016} (\bibinfo {year} {2010})},\ \Eprint {http://arxiv.org/abs/1004.4697} {arXiv:1004.4697 [gr-qc]} \BibitemShut {NoStop}%
\bibitem [{\citenamefont {Healy}\ \emph {et~al.}(2017)\citenamefont {Healy}, \citenamefont {Lousto}, \citenamefont {Nakano},\ and\ \citenamefont {Zlochower}}]{Healy:2017zqj}%
  \BibitemOpen
  \bibfield  {author} {\bibinfo {author} {\bibfnamefont {James}\ \bibnamefont {Healy}}, \bibinfo {author} {\bibfnamefont {Carlos~O.}\ \bibnamefont {Lousto}}, \bibinfo {author} {\bibfnamefont {Hiroyuki}\ \bibnamefont {Nakano}}, \ and\ \bibinfo {author} {\bibfnamefont {Yosef}\ \bibnamefont {Zlochower}},\ }\bibfield  {title} {\enquote {\bibinfo {title} {{Post-Newtonian Quasicircular Initial Orbits for Numerical Relativity}},}\ }\href {\doibase 10.1088/1361-6382/aa7929} {\bibfield  {journal} {\bibinfo  {journal} {Class. Quant. Grav.}\ }\textbf {\bibinfo {volume} {34}},\ \bibinfo {pages} {145011} (\bibinfo {year} {2017})},\ \bibinfo {note} {[Erratum: Class.Quant.Grav. 40, 249502 (2023)]},\ \Eprint {http://arxiv.org/abs/1702.00872} {arXiv:1702.00872 [gr-qc]} \BibitemShut {NoStop}%
\bibitem [{\citenamefont {Buonanno}\ \emph {et~al.}(2007)\citenamefont {Buonanno}, \citenamefont {Cook},\ and\ \citenamefont {Pretorius}}]{Buonanno:2006ui}%
  \BibitemOpen
  \bibfield  {author} {\bibinfo {author} {\bibfnamefont {Alessandra}\ \bibnamefont {Buonanno}}, \bibinfo {author} {\bibfnamefont {Gregory~B.}\ \bibnamefont {Cook}}, \ and\ \bibinfo {author} {\bibfnamefont {Frans}\ \bibnamefont {Pretorius}},\ }\bibfield  {title} {\enquote {\bibinfo {title} {{Inspiral, merger and ring-down of equal-mass black-hole binaries}},}\ }\href {\doibase 10.1103/PhysRevD.75.124018} {\bibfield  {journal} {\bibinfo  {journal} {Phys. Rev. D}\ }\textbf {\bibinfo {volume} {75}},\ \bibinfo {pages} {124018} (\bibinfo {year} {2007})},\ \Eprint {http://arxiv.org/abs/gr-qc/0610122} {arXiv:gr-qc/0610122} \BibitemShut {NoStop}%
\bibitem [{\citenamefont {Husa}\ \emph {et~al.}(2008)\citenamefont {Husa}, \citenamefont {Hannam}, \citenamefont {Gonzalez}, \citenamefont {Sperhake},\ and\ \citenamefont {Bruegmann}}]{Husa:2007rh}%
  \BibitemOpen
  \bibfield  {author} {\bibinfo {author} {\bibfnamefont {Sascha}\ \bibnamefont {Husa}}, \bibinfo {author} {\bibfnamefont {Mark}\ \bibnamefont {Hannam}}, \bibinfo {author} {\bibfnamefont {Jose~A.}\ \bibnamefont {Gonzalez}}, \bibinfo {author} {\bibfnamefont {Ulrich}\ \bibnamefont {Sperhake}}, \ and\ \bibinfo {author} {\bibfnamefont {Bernd}\ \bibnamefont {Bruegmann}},\ }\bibfield  {title} {\enquote {\bibinfo {title} {{Reducing eccentricity in black-hole binary evolutions with initial parameters from post-Newtonian inspiral}},}\ }\href {\doibase 10.1103/PhysRevD.77.044037} {\bibfield  {journal} {\bibinfo  {journal} {Phys. Rev. D}\ }\textbf {\bibinfo {volume} {77}},\ \bibinfo {pages} {044037} (\bibinfo {year} {2008})},\ \Eprint {http://arxiv.org/abs/0706.0904} {arXiv:0706.0904 [gr-qc]} \BibitemShut {NoStop}%
\bibitem [{\citenamefont {Ramos-Buades}\ \emph {et~al.}(2019)\citenamefont {Ramos-Buades}, \citenamefont {Husa},\ and\ \citenamefont {Pratten}}]{Ramos-Buades:2018azo}%
  \BibitemOpen
  \bibfield  {author} {\bibinfo {author} {\bibfnamefont {Antoni}\ \bibnamefont {Ramos-Buades}}, \bibinfo {author} {\bibfnamefont {Sascha}\ \bibnamefont {Husa}}, \ and\ \bibinfo {author} {\bibfnamefont {Geraint}\ \bibnamefont {Pratten}},\ }\bibfield  {title} {\enquote {\bibinfo {title} {{Simple procedures to reduce eccentricity of binary black hole simulations}},}\ }\href {\doibase 10.1103/PhysRevD.99.023003} {\bibfield  {journal} {\bibinfo  {journal} {Phys. Rev. D}\ }\textbf {\bibinfo {volume} {99}},\ \bibinfo {pages} {023003} (\bibinfo {year} {2019})},\ \Eprint {http://arxiv.org/abs/1810.00036} {arXiv:1810.00036 [gr-qc]} \BibitemShut {NoStop}%
\bibitem [{\citenamefont {Ramos-Buades}\ \emph {et~al.}(2020)\citenamefont {Ramos-Buades}, \citenamefont {Husa}, \citenamefont {Pratten}, \citenamefont {Estell\'es}, \citenamefont {Garc\'\i{}a-Quir\'os}, \citenamefont {Mateu-Lucena}, \citenamefont {Colleoni},\ and\ \citenamefont {Jaume}}]{Ramos-Buades:2019uvh}%
  \BibitemOpen
  \bibfield  {author} {\bibinfo {author} {\bibfnamefont {Antoni}\ \bibnamefont {Ramos-Buades}}, \bibinfo {author} {\bibfnamefont {Sascha}\ \bibnamefont {Husa}}, \bibinfo {author} {\bibfnamefont {Geraint}\ \bibnamefont {Pratten}}, \bibinfo {author} {\bibfnamefont {H\'ector}\ \bibnamefont {Estell\'es}}, \bibinfo {author} {\bibfnamefont {Cecilio}\ \bibnamefont {Garc\'\i{}a-Quir\'os}}, \bibinfo {author} {\bibfnamefont {Maite}\ \bibnamefont {Mateu-Lucena}}, \bibinfo {author} {\bibfnamefont {Marta}\ \bibnamefont {Colleoni}}, \ and\ \bibinfo {author} {\bibfnamefont {Rafel}\ \bibnamefont {Jaume}},\ }\bibfield  {title} {\enquote {\bibinfo {title} {{First survey of spinning eccentric black hole mergers: Numerical relativity simulations, hybrid waveforms, and parameter estimation}},}\ }\href {\doibase 10.1103/PhysRevD.101.083015} {\bibfield  {journal} {\bibinfo  {journal} {Phys. Rev. D}\ }\textbf {\bibinfo {volume} {101}},\ \bibinfo {pages} {083015} (\bibinfo {year} {2020})},\ \Eprint {http://arxiv.org/abs/1909.11011}
  {arXiv:1909.11011 [gr-qc]} \BibitemShut {NoStop}%
\bibitem [{\citenamefont {Purrer}\ \emph {et~al.}(2012)\citenamefont {Purrer}, \citenamefont {Husa},\ and\ \citenamefont {Hannam}}]{Purrer:2012wy}%
  \BibitemOpen
  \bibfield  {author} {\bibinfo {author} {\bibfnamefont {Michael}\ \bibnamefont {Purrer}}, \bibinfo {author} {\bibfnamefont {Sascha}\ \bibnamefont {Husa}}, \ and\ \bibinfo {author} {\bibfnamefont {Mark}\ \bibnamefont {Hannam}},\ }\bibfield  {title} {\enquote {\bibinfo {title} {{An Efficient iterative method to reduce eccentricity in numerical-relativity simulations of compact binary inspiral}},}\ }\href {\doibase 10.1103/PhysRevD.85.124051} {\bibfield  {journal} {\bibinfo  {journal} {Phys. Rev. D}\ }\textbf {\bibinfo {volume} {85}},\ \bibinfo {pages} {124051} (\bibinfo {year} {2012})},\ \Eprint {http://arxiv.org/abs/1203.4258} {arXiv:1203.4258 [gr-qc]} \BibitemShut {NoStop}%
\bibitem [{\citenamefont {Bonino}\ \emph {et~al.}(2024)\citenamefont {Bonino}, \citenamefont {Schmidt},\ and\ \citenamefont {Pratten}}]{Bonino:2024xrv}%
  \BibitemOpen
  \bibfield  {author} {\bibinfo {author} {\bibfnamefont {Alice}\ \bibnamefont {Bonino}}, \bibinfo {author} {\bibfnamefont {Patricia}\ \bibnamefont {Schmidt}}, \ and\ \bibinfo {author} {\bibfnamefont {Geraint}\ \bibnamefont {Pratten}},\ }\bibfield  {title} {\enquote {\bibinfo {title} {{Mapping eccentricity evolutions between numerical relativity and effective-one-body gravitational waveforms}},}\ }\href@noop {} {\  (\bibinfo {year} {2024})},\ \Eprint {http://arxiv.org/abs/2404.18875} {arXiv:2404.18875 [gr-qc]} \BibitemShut {NoStop}%
\bibitem [{\citenamefont {Ramos-Buades}\ \emph {et~al.}(2022{\natexlab{a}})\citenamefont {Ramos-Buades}, \citenamefont {van~de Meent}, \citenamefont {Pfeiffer}, \citenamefont {R\"uter}, \citenamefont {Scheel}, \citenamefont {Boyle},\ and\ \citenamefont {Kidder}}]{Ramos-Buades:2022lgf}%
  \BibitemOpen
  \bibfield  {author} {\bibinfo {author} {\bibfnamefont {Antoni}\ \bibnamefont {Ramos-Buades}}, \bibinfo {author} {\bibfnamefont {Maarten}\ \bibnamefont {van~de Meent}}, \bibinfo {author} {\bibfnamefont {Harald~P.}\ \bibnamefont {Pfeiffer}}, \bibinfo {author} {\bibfnamefont {Hannes~R.}\ \bibnamefont {R\"uter}}, \bibinfo {author} {\bibfnamefont {Mark~A.}\ \bibnamefont {Scheel}}, \bibinfo {author} {\bibfnamefont {Michael}\ \bibnamefont {Boyle}}, \ and\ \bibinfo {author} {\bibfnamefont {Lawrence~E.}\ \bibnamefont {Kidder}},\ }\bibfield  {title} {\enquote {\bibinfo {title} {{Eccentric binary black holes: Comparing numerical relativity and small mass-ratio perturbation theory}},}\ }\href {\doibase 10.1103/PhysRevD.106.124040} {\bibfield  {journal} {\bibinfo  {journal} {Phys. Rev. D}\ }\textbf {\bibinfo {volume} {106}},\ \bibinfo {pages} {124040} (\bibinfo {year} {2022}{\natexlab{a}})},\ \Eprint {http://arxiv.org/abs/2209.03390} {arXiv:2209.03390 [gr-qc]} \BibitemShut {NoStop}%
\bibitem [{\citenamefont {Arun}\ \emph {et~al.}(2009)\citenamefont {Arun}, \citenamefont {Blanchet}, \citenamefont {Iyer},\ and\ \citenamefont {Sinha}}]{Arun:2009mc}%
  \BibitemOpen
  \bibfield  {author} {\bibinfo {author} {\bibfnamefont {K.~G.}\ \bibnamefont {Arun}}, \bibinfo {author} {\bibfnamefont {Luc}\ \bibnamefont {Blanchet}}, \bibinfo {author} {\bibfnamefont {Bala~R.}\ \bibnamefont {Iyer}}, \ and\ \bibinfo {author} {\bibfnamefont {Siddhartha}\ \bibnamefont {Sinha}},\ }\bibfield  {title} {\enquote {\bibinfo {title} {{Third post-Newtonian angular momentum flux and the secular evolution of orbital elements for inspiralling compact binaries in quasi-elliptical orbits}},}\ }\href {\doibase 10.1103/PhysRevD.80.124018} {\bibfield  {journal} {\bibinfo  {journal} {Phys. Rev. D}\ }\textbf {\bibinfo {volume} {80}},\ \bibinfo {pages} {124018} (\bibinfo {year} {2009})},\ \Eprint {http://arxiv.org/abs/0908.3854} {arXiv:0908.3854 [gr-qc]} \BibitemShut {NoStop}%
\bibitem [{\citenamefont {Tanay}\ \emph {et~al.}(2016)\citenamefont {Tanay}, \citenamefont {Haney},\ and\ \citenamefont {Gopakumar}}]{Tanay:2016zog}%
  \BibitemOpen
  \bibfield  {author} {\bibinfo {author} {\bibfnamefont {Sashwat}\ \bibnamefont {Tanay}}, \bibinfo {author} {\bibfnamefont {Maria}\ \bibnamefont {Haney}}, \ and\ \bibinfo {author} {\bibfnamefont {Achamveedu}\ \bibnamefont {Gopakumar}},\ }\bibfield  {title} {\enquote {\bibinfo {title} {{Frequency and time domain inspiral templates for comparable mass compact binaries in eccentric orbits}},}\ }\href {\doibase 10.1103/PhysRevD.93.064031} {\bibfield  {journal} {\bibinfo  {journal} {Phys. Rev. D}\ }\textbf {\bibinfo {volume} {93}},\ \bibinfo {pages} {064031} (\bibinfo {year} {2016})},\ \Eprint {http://arxiv.org/abs/1602.03081} {arXiv:1602.03081 [gr-qc]} \BibitemShut {NoStop}%
\bibitem [{\citenamefont {Paul}\ and\ \citenamefont {Mishra}(2023)}]{Paul:2022xfy}%
  \BibitemOpen
  \bibfield  {author} {\bibinfo {author} {\bibfnamefont {Kaushik}\ \bibnamefont {Paul}}\ and\ \bibinfo {author} {\bibfnamefont {Chandra~Kant}\ \bibnamefont {Mishra}},\ }\bibfield  {title} {\enquote {\bibinfo {title} {{Spin effects in spherical harmonic modes of gravitational waves from eccentric compact binary inspirals}},}\ }\href {\doibase 10.1103/PhysRevD.108.024023} {\bibfield  {journal} {\bibinfo  {journal} {Phys. Rev. D}\ }\textbf {\bibinfo {volume} {108}},\ \bibinfo {pages} {024023} (\bibinfo {year} {2023})},\ \Eprint {http://arxiv.org/abs/2211.04155} {arXiv:2211.04155 [gr-qc]} \BibitemShut {NoStop}%
\bibitem [{\citenamefont {Henry}\ and\ \citenamefont {Khalil}(2023)}]{Henry:2023tka}%
  \BibitemOpen
  \bibfield  {author} {\bibinfo {author} {\bibfnamefont {Quentin}\ \bibnamefont {Henry}}\ and\ \bibinfo {author} {\bibfnamefont {Mohammed}\ \bibnamefont {Khalil}},\ }\bibfield  {title} {\enquote {\bibinfo {title} {{Spin effects in gravitational waveforms and fluxes for binaries on eccentric orbits to the third post-Newtonian order}},}\ }\href {\doibase 10.1103/PhysRevD.108.104016} {\bibfield  {journal} {\bibinfo  {journal} {Phys. Rev. D}\ }\textbf {\bibinfo {volume} {108}},\ \bibinfo {pages} {104016} (\bibinfo {year} {2023})},\ \Eprint {http://arxiv.org/abs/2308.13606} {arXiv:2308.13606 [gr-qc]} \BibitemShut {NoStop}%
\bibitem [{\citenamefont {Tiwari}\ \emph {et~al.}(2019)\citenamefont {Tiwari}, \citenamefont {Achamveedu}, \citenamefont {Haney},\ and\ \citenamefont {Hemantakumar}}]{Tiwari:2019jtz}%
  \BibitemOpen
  \bibfield  {author} {\bibinfo {author} {\bibfnamefont {Srishti}\ \bibnamefont {Tiwari}}, \bibinfo {author} {\bibfnamefont {Gopakumar}\ \bibnamefont {Achamveedu}}, \bibinfo {author} {\bibfnamefont {Maria}\ \bibnamefont {Haney}}, \ and\ \bibinfo {author} {\bibfnamefont {Phurailatapam}\ \bibnamefont {Hemantakumar}},\ }\bibfield  {title} {\enquote {\bibinfo {title} {{Ready-to-use Fourier domain templates for compact binaries inspiraling along moderately eccentric orbits}},}\ }\href {\doibase 10.1103/PhysRevD.99.124008} {\bibfield  {journal} {\bibinfo  {journal} {Phys. Rev. D}\ }\textbf {\bibinfo {volume} {99}},\ \bibinfo {pages} {124008} (\bibinfo {year} {2019})},\ \Eprint {http://arxiv.org/abs/1905.07956} {arXiv:1905.07956 [gr-qc]} \BibitemShut {NoStop}%
\bibitem [{\citenamefont {Huerta}\ \emph {et~al.}(2014)\citenamefont {Huerta}, \citenamefont {Kumar}, \citenamefont {McWilliams}, \citenamefont {O'Shaughnessy},\ and\ \citenamefont {Yunes}}]{Huerta:2014eca}%
  \BibitemOpen
  \bibfield  {author} {\bibinfo {author} {\bibfnamefont {E.~A.}\ \bibnamefont {Huerta}}, \bibinfo {author} {\bibfnamefont {Prayush}\ \bibnamefont {Kumar}}, \bibinfo {author} {\bibfnamefont {Sean~T.}\ \bibnamefont {McWilliams}}, \bibinfo {author} {\bibfnamefont {Richard}\ \bibnamefont {O'Shaughnessy}}, \ and\ \bibinfo {author} {\bibfnamefont {Nicol\'as}\ \bibnamefont {Yunes}},\ }\bibfield  {title} {\enquote {\bibinfo {title} {{Accurate and efficient waveforms for compact binaries on eccentric orbits}},}\ }\href {\doibase 10.1103/PhysRevD.90.084016} {\bibfield  {journal} {\bibinfo  {journal} {Phys. Rev. D}\ }\textbf {\bibinfo {volume} {90}},\ \bibinfo {pages} {084016} (\bibinfo {year} {2014})},\ \Eprint {http://arxiv.org/abs/1408.3406} {arXiv:1408.3406 [gr-qc]} \BibitemShut {NoStop}%
\bibitem [{\citenamefont {Moore}\ \emph {et~al.}(2016)\citenamefont {Moore}, \citenamefont {Favata}, \citenamefont {Arun},\ and\ \citenamefont {Mishra}}]{Moore:2016qxz}%
  \BibitemOpen
  \bibfield  {author} {\bibinfo {author} {\bibfnamefont {Blake}\ \bibnamefont {Moore}}, \bibinfo {author} {\bibfnamefont {Marc}\ \bibnamefont {Favata}}, \bibinfo {author} {\bibfnamefont {K.~G.}\ \bibnamefont {Arun}}, \ and\ \bibinfo {author} {\bibfnamefont {Chandra~Kant}\ \bibnamefont {Mishra}},\ }\bibfield  {title} {\enquote {\bibinfo {title} {{Gravitational-wave phasing for low-eccentricity inspiralling compact binaries to 3PN order}},}\ }\href {\doibase 10.1103/PhysRevD.93.124061} {\bibfield  {journal} {\bibinfo  {journal} {Phys. Rev. D}\ }\textbf {\bibinfo {volume} {93}},\ \bibinfo {pages} {124061} (\bibinfo {year} {2016})},\ \Eprint {http://arxiv.org/abs/1605.00304} {arXiv:1605.00304 [gr-qc]} \BibitemShut {NoStop}%
\bibitem [{\citenamefont {Damour}\ \emph {et~al.}(2004)\citenamefont {Damour}, \citenamefont {Gopakumar},\ and\ \citenamefont {Iyer}}]{Damour:2004bz}%
  \BibitemOpen
  \bibfield  {author} {\bibinfo {author} {\bibfnamefont {Thibault}\ \bibnamefont {Damour}}, \bibinfo {author} {\bibfnamefont {Achamveedu}\ \bibnamefont {Gopakumar}}, \ and\ \bibinfo {author} {\bibfnamefont {Bala~R.}\ \bibnamefont {Iyer}},\ }\bibfield  {title} {\enquote {\bibinfo {title} {{Phasing of gravitational waves from inspiralling eccentric binaries}},}\ }\href {\doibase 10.1103/PhysRevD.70.064028} {\bibfield  {journal} {\bibinfo  {journal} {Phys. Rev. D}\ }\textbf {\bibinfo {volume} {70}},\ \bibinfo {pages} {064028} (\bibinfo {year} {2004})},\ \Eprint {http://arxiv.org/abs/gr-qc/0404128} {arXiv:gr-qc/0404128} \BibitemShut {NoStop}%
\bibitem [{\citenamefont {Konigsdorffer}\ and\ \citenamefont {Gopakumar}(2006)}]{Konigsdorffer:2006zt}%
  \BibitemOpen
  \bibfield  {author} {\bibinfo {author} {\bibfnamefont {Christian}\ \bibnamefont {Konigsdorffer}}\ and\ \bibinfo {author} {\bibfnamefont {Achamveedu}\ \bibnamefont {Gopakumar}},\ }\bibfield  {title} {\enquote {\bibinfo {title} {{Phasing of gravitational waves from inspiralling eccentric binaries at the third-and-a-half post-Newtonian order}},}\ }\href {\doibase 10.1103/PhysRevD.73.124012} {\bibfield  {journal} {\bibinfo  {journal} {Phys. Rev. D}\ }\textbf {\bibinfo {volume} {73}},\ \bibinfo {pages} {124012} (\bibinfo {year} {2006})},\ \Eprint {http://arxiv.org/abs/gr-qc/0603056} {arXiv:gr-qc/0603056} \BibitemShut {NoStop}%
\bibitem [{\citenamefont {Memmesheimer}\ \emph {et~al.}(2004)\citenamefont {Memmesheimer}, \citenamefont {Gopakumar},\ and\ \citenamefont {Schaefer}}]{Memmesheimer:2004cv}%
  \BibitemOpen
  \bibfield  {author} {\bibinfo {author} {\bibfnamefont {Raoul-Martin}\ \bibnamefont {Memmesheimer}}, \bibinfo {author} {\bibfnamefont {Achamveedu}\ \bibnamefont {Gopakumar}}, \ and\ \bibinfo {author} {\bibfnamefont {Gerhard}\ \bibnamefont {Schaefer}},\ }\bibfield  {title} {\enquote {\bibinfo {title} {{Third post-Newtonian accurate generalized quasi-Keplerian parametrization for compact binaries in eccentric orbits}},}\ }\href {\doibase 10.1103/PhysRevD.70.104011} {\bibfield  {journal} {\bibinfo  {journal} {Phys. Rev. D}\ }\textbf {\bibinfo {volume} {70}},\ \bibinfo {pages} {104011} (\bibinfo {year} {2004})},\ \Eprint {http://arxiv.org/abs/gr-qc/0407049} {arXiv:gr-qc/0407049} \BibitemShut {NoStop}%
\bibitem [{\citenamefont {Hinder}\ \emph {et~al.}(2018)\citenamefont {Hinder}, \citenamefont {Kidder},\ and\ \citenamefont {Pfeiffer}}]{Hinder:2017sxy}%
  \BibitemOpen
  \bibfield  {author} {\bibinfo {author} {\bibfnamefont {Ian}\ \bibnamefont {Hinder}}, \bibinfo {author} {\bibfnamefont {Lawrence~E.}\ \bibnamefont {Kidder}}, \ and\ \bibinfo {author} {\bibfnamefont {Harald~P.}\ \bibnamefont {Pfeiffer}},\ }\bibfield  {title} {\enquote {\bibinfo {title} {{Eccentric binary black hole inspiral-merger-ringdown gravitational waveform model from numerical relativity and post-Newtonian theory}},}\ }\href {\doibase 10.1103/PhysRevD.98.044015} {\bibfield  {journal} {\bibinfo  {journal} {Phys. Rev. D}\ }\textbf {\bibinfo {volume} {98}},\ \bibinfo {pages} {044015} (\bibinfo {year} {2018})},\ \Eprint {http://arxiv.org/abs/1709.02007} {arXiv:1709.02007 [gr-qc]} \BibitemShut {NoStop}%
\bibitem [{\citenamefont {Cho}\ \emph {et~al.}(2022)\citenamefont {Cho}, \citenamefont {Tanay}, \citenamefont {Gopakumar},\ and\ \citenamefont {Lee}}]{Cho:2021oai}%
  \BibitemOpen
  \bibfield  {author} {\bibinfo {author} {\bibfnamefont {Gihyuk}\ \bibnamefont {Cho}}, \bibinfo {author} {\bibfnamefont {Sashwat}\ \bibnamefont {Tanay}}, \bibinfo {author} {\bibfnamefont {Achamveedu}\ \bibnamefont {Gopakumar}}, \ and\ \bibinfo {author} {\bibfnamefont {Hyung~Mok}\ \bibnamefont {Lee}},\ }\bibfield  {title} {\enquote {\bibinfo {title} {{Generalized quasi-Keplerian solution for eccentric, nonspinning compact binaries at 4PN order and the associated inspiral-merger-ringdown waveform}},}\ }\href {\doibase 10.1103/PhysRevD.105.064010} {\bibfield  {journal} {\bibinfo  {journal} {Phys. Rev. D}\ }\textbf {\bibinfo {volume} {105}},\ \bibinfo {pages} {064010} (\bibinfo {year} {2022})},\ \Eprint {http://arxiv.org/abs/2110.09608} {arXiv:2110.09608 [gr-qc]} \BibitemShut {NoStop}%
\bibitem [{\citenamefont {Chattaraj}\ \emph {et~al.}(2022)\citenamefont {Chattaraj}, \citenamefont {RoyChowdhury}, \citenamefont {Divyajyoti}, \citenamefont {Mishra},\ and\ \citenamefont {Gupta}}]{Chattaraj:2022tay}%
  \BibitemOpen
  \bibfield  {author} {\bibinfo {author} {\bibfnamefont {Abhishek}\ \bibnamefont {Chattaraj}}, \bibinfo {author} {\bibfnamefont {Tamal}\ \bibnamefont {RoyChowdhury}}, \bibinfo {author} {\bibnamefont {Divyajyoti}}, \bibinfo {author} {\bibfnamefont {Chandra~Kant}\ \bibnamefont {Mishra}}, \ and\ \bibinfo {author} {\bibfnamefont {Anshu}\ \bibnamefont {Gupta}},\ }\bibfield  {title} {\enquote {\bibinfo {title} {{High accuracy post-Newtonian and numerical relativity comparisons involving higher modes for eccentric binary black holes and a dominant mode eccentric inspiral-merger-ringdown model}},}\ }\href {\doibase 10.1103/PhysRevD.106.124008} {\bibfield  {journal} {\bibinfo  {journal} {Phys. Rev. D}\ }\textbf {\bibinfo {volume} {106}},\ \bibinfo {pages} {124008} (\bibinfo {year} {2022})},\ \Eprint {http://arxiv.org/abs/2204.02377} {arXiv:2204.02377 [gr-qc]} \BibitemShut {NoStop}%
\bibitem [{\citenamefont {Hinderer}\ and\ \citenamefont {Babak}(2017)}]{Hinderer:2017jcs}%
  \BibitemOpen
  \bibfield  {author} {\bibinfo {author} {\bibfnamefont {Tanja}\ \bibnamefont {Hinderer}}\ and\ \bibinfo {author} {\bibfnamefont {Stanislav}\ \bibnamefont {Babak}},\ }\bibfield  {title} {\enquote {\bibinfo {title} {{Foundations of an effective-one-body model for coalescing binaries on eccentric orbits}},}\ }\href {\doibase 10.1103/PhysRevD.96.104048} {\bibfield  {journal} {\bibinfo  {journal} {Phys. Rev. D}\ }\textbf {\bibinfo {volume} {96}},\ \bibinfo {pages} {104048} (\bibinfo {year} {2017})},\ \Eprint {http://arxiv.org/abs/1707.08426} {arXiv:1707.08426 [gr-qc]} \BibitemShut {NoStop}%
\bibitem [{\citenamefont {Cao}\ and\ \citenamefont {Han}(2017)}]{Cao:2017ndf}%
  \BibitemOpen
  \bibfield  {author} {\bibinfo {author} {\bibfnamefont {Zhoujian}\ \bibnamefont {Cao}}\ and\ \bibinfo {author} {\bibfnamefont {Wen-Biao}\ \bibnamefont {Han}},\ }\bibfield  {title} {\enquote {\bibinfo {title} {{Waveform model for an eccentric binary black hole based on the effective-one-body-numerical-relativity formalism}},}\ }\href {\doibase 10.1103/PhysRevD.96.044028} {\bibfield  {journal} {\bibinfo  {journal} {Phys. Rev. D}\ }\textbf {\bibinfo {volume} {96}},\ \bibinfo {pages} {044028} (\bibinfo {year} {2017})},\ \Eprint {http://arxiv.org/abs/1708.00166} {arXiv:1708.00166 [gr-qc]} \BibitemShut {NoStop}%
\bibitem [{\citenamefont {Chiaramello}\ and\ \citenamefont {Nagar}(2020)}]{Chiaramello:2020ehz}%
  \BibitemOpen
  \bibfield  {author} {\bibinfo {author} {\bibfnamefont {Danilo}\ \bibnamefont {Chiaramello}}\ and\ \bibinfo {author} {\bibfnamefont {Alessandro}\ \bibnamefont {Nagar}},\ }\bibfield  {title} {\enquote {\bibinfo {title} {{Faithful analytical effective-one-body waveform model for spin-aligned, moderately eccentric, coalescing black hole binaries}},}\ }\href {\doibase 10.1103/PhysRevD.101.101501} {\bibfield  {journal} {\bibinfo  {journal} {Phys. Rev. D}\ }\textbf {\bibinfo {volume} {101}},\ \bibinfo {pages} {101501} (\bibinfo {year} {2020})},\ \Eprint {http://arxiv.org/abs/2001.11736} {arXiv:2001.11736 [gr-qc]} \BibitemShut {NoStop}%
\bibitem [{\citenamefont {Albanesi}\ \emph {et~al.}(2023)\citenamefont {Albanesi}, \citenamefont {Bernuzzi}, \citenamefont {Damour}, \citenamefont {Nagar},\ and\ \citenamefont {Placidi}}]{Albanesi:2023bgi}%
  \BibitemOpen
  \bibfield  {author} {\bibinfo {author} {\bibfnamefont {Simone}\ \bibnamefont {Albanesi}}, \bibinfo {author} {\bibfnamefont {Sebastiano}\ \bibnamefont {Bernuzzi}}, \bibinfo {author} {\bibfnamefont {Thibault}\ \bibnamefont {Damour}}, \bibinfo {author} {\bibfnamefont {Alessandro}\ \bibnamefont {Nagar}}, \ and\ \bibinfo {author} {\bibfnamefont {Andrea}\ \bibnamefont {Placidi}},\ }\bibfield  {title} {\enquote {\bibinfo {title} {{Faithful effective-one-body waveform of small-mass-ratio coalescing black hole binaries: The eccentric, nonspinning case}},}\ }\href {\doibase 10.1103/PhysRevD.108.084037} {\bibfield  {journal} {\bibinfo  {journal} {Phys. Rev. D}\ }\textbf {\bibinfo {volume} {108}},\ \bibinfo {pages} {084037} (\bibinfo {year} {2023})},\ \Eprint {http://arxiv.org/abs/2305.19336} {arXiv:2305.19336 [gr-qc]} \BibitemShut {NoStop}%
\bibitem [{\citenamefont {Albanesi}\ \emph {et~al.}(2022)\citenamefont {Albanesi}, \citenamefont {Placidi}, \citenamefont {Nagar}, \citenamefont {Orselli},\ and\ \citenamefont {Bernuzzi}}]{Albanesi:2022xge}%
  \BibitemOpen
  \bibfield  {author} {\bibinfo {author} {\bibfnamefont {Simone}\ \bibnamefont {Albanesi}}, \bibinfo {author} {\bibfnamefont {Andrea}\ \bibnamefont {Placidi}}, \bibinfo {author} {\bibfnamefont {Alessandro}\ \bibnamefont {Nagar}}, \bibinfo {author} {\bibfnamefont {Marta}\ \bibnamefont {Orselli}}, \ and\ \bibinfo {author} {\bibfnamefont {Sebastiano}\ \bibnamefont {Bernuzzi}},\ }\bibfield  {title} {\enquote {\bibinfo {title} {{New avenue for accurate analytical waveforms and fluxes for eccentric compact binaries}},}\ }\href {\doibase 10.1103/PhysRevD.105.L121503} {\bibfield  {journal} {\bibinfo  {journal} {Phys. Rev. D}\ }\textbf {\bibinfo {volume} {105}},\ \bibinfo {pages} {L121503} (\bibinfo {year} {2022})},\ \Eprint {http://arxiv.org/abs/2203.16286} {arXiv:2203.16286 [gr-qc]} \BibitemShut {NoStop}%
\bibitem [{\citenamefont {Riemenschneider}\ \emph {et~al.}(2021)\citenamefont {Riemenschneider}, \citenamefont {Rettegno}, \citenamefont {Breschi}, \citenamefont {Albertini}, \citenamefont {Gamba}, \citenamefont {Bernuzzi},\ and\ \citenamefont {Nagar}}]{Riemenschneider:2021ppj}%
  \BibitemOpen
  \bibfield  {author} {\bibinfo {author} {\bibfnamefont {Gunnar}\ \bibnamefont {Riemenschneider}}, \bibinfo {author} {\bibfnamefont {Piero}\ \bibnamefont {Rettegno}}, \bibinfo {author} {\bibfnamefont {Matteo}\ \bibnamefont {Breschi}}, \bibinfo {author} {\bibfnamefont {Angelica}\ \bibnamefont {Albertini}}, \bibinfo {author} {\bibfnamefont {Rossella}\ \bibnamefont {Gamba}}, \bibinfo {author} {\bibfnamefont {Sebastiano}\ \bibnamefont {Bernuzzi}}, \ and\ \bibinfo {author} {\bibfnamefont {Alessandro}\ \bibnamefont {Nagar}},\ }\bibfield  {title} {\enquote {\bibinfo {title} {{Assessment of consistent next-to-quasicircular corrections and postadiabatic approximation in effective-one-body multipolar waveforms for binary black hole coalescences}},}\ }\href {\doibase 10.1103/PhysRevD.104.104045} {\bibfield  {journal} {\bibinfo  {journal} {Phys. Rev. D}\ }\textbf {\bibinfo {volume} {104}},\ \bibinfo {pages} {104045} (\bibinfo {year} {2021})},\ \Eprint {http://arxiv.org/abs/2104.07533} {arXiv:2104.07533 [gr-qc]}
  \BibitemShut {NoStop}%
\bibitem [{\citenamefont {Ramos-Buades}\ \emph {et~al.}(2022{\natexlab{b}})\citenamefont {Ramos-Buades}, \citenamefont {Buonanno}, \citenamefont {Khalil},\ and\ \citenamefont {Ossokine}}]{Ramos-Buades:2021adz}%
  \BibitemOpen
  \bibfield  {author} {\bibinfo {author} {\bibfnamefont {Antoni}\ \bibnamefont {Ramos-Buades}}, \bibinfo {author} {\bibfnamefont {Alessandra}\ \bibnamefont {Buonanno}}, \bibinfo {author} {\bibfnamefont {Mohammed}\ \bibnamefont {Khalil}}, \ and\ \bibinfo {author} {\bibfnamefont {Serguei}\ \bibnamefont {Ossokine}},\ }\bibfield  {title} {\enquote {\bibinfo {title} {{Effective-one-body multipolar waveforms for eccentric binary black holes with nonprecessing spins}},}\ }\href {\doibase 10.1103/PhysRevD.105.044035} {\bibfield  {journal} {\bibinfo  {journal} {Phys. Rev. D}\ }\textbf {\bibinfo {volume} {105}},\ \bibinfo {pages} {044035} (\bibinfo {year} {2022}{\natexlab{b}})},\ \Eprint {http://arxiv.org/abs/2112.06952} {arXiv:2112.06952 [gr-qc]} \BibitemShut {NoStop}%
\bibitem [{\citenamefont {Liu}\ \emph {et~al.}(2023)\citenamefont {Liu}, \citenamefont {Cao},\ and\ \citenamefont {Zhu}}]{Liu:2023ldr}%
  \BibitemOpen
  \bibfield  {author} {\bibinfo {author} {\bibfnamefont {Xiaolin}\ \bibnamefont {Liu}}, \bibinfo {author} {\bibfnamefont {Zhoujian}\ \bibnamefont {Cao}}, \ and\ \bibinfo {author} {\bibfnamefont {Zong-Hong}\ \bibnamefont {Zhu}},\ }\bibfield  {title} {\enquote {\bibinfo {title} {{Effective-One-Body Numerical-Relativity waveform model for Eccentric spin-precessing binary black hole coalescence}},}\ }\href@noop {} {\  (\bibinfo {year} {2023})},\ \Eprint {http://arxiv.org/abs/2310.04552} {arXiv:2310.04552 [gr-qc]} \BibitemShut {NoStop}%
\bibitem [{\citenamefont {Huerta}\ \emph {et~al.}(2017)\citenamefont {Huerta} \emph {et~al.}}]{Huerta:2016rwp}%
  \BibitemOpen
  \bibfield  {author} {\bibinfo {author} {\bibfnamefont {E.~A.}\ \bibnamefont {Huerta}} \emph {et~al.},\ }\bibfield  {title} {\enquote {\bibinfo {title} {{Complete waveform model for compact binaries on eccentric orbits}},}\ }\href {\doibase 10.1103/PhysRevD.95.024038} {\bibfield  {journal} {\bibinfo  {journal} {Phys. Rev. D}\ }\textbf {\bibinfo {volume} {95}},\ \bibinfo {pages} {024038} (\bibinfo {year} {2017})},\ \Eprint {http://arxiv.org/abs/1609.05933} {arXiv:1609.05933 [gr-qc]} \BibitemShut {NoStop}%
\bibitem [{\citenamefont {Huerta}\ \emph {et~al.}(2018)\citenamefont {Huerta} \emph {et~al.}}]{Huerta:2017kez}%
  \BibitemOpen
  \bibfield  {author} {\bibinfo {author} {\bibfnamefont {E.~A.}\ \bibnamefont {Huerta}} \emph {et~al.},\ }\bibfield  {title} {\enquote {\bibinfo {title} {{Eccentric, nonspinning, inspiral, Gaussian-process merger approximant for the detection and characterization of eccentric binary black hole mergers}},}\ }\href {\doibase 10.1103/PhysRevD.97.024031} {\bibfield  {journal} {\bibinfo  {journal} {Phys. Rev. D}\ }\textbf {\bibinfo {volume} {97}},\ \bibinfo {pages} {024031} (\bibinfo {year} {2018})},\ \Eprint {http://arxiv.org/abs/1711.06276} {arXiv:1711.06276 [gr-qc]} \BibitemShut {NoStop}%
\bibitem [{\citenamefont {Joshi}\ \emph {et~al.}(2023)\citenamefont {Joshi}, \citenamefont {Rosofsky}, \citenamefont {Haas},\ and\ \citenamefont {Huerta}}]{Joshi:2022ocr}%
  \BibitemOpen
  \bibfield  {author} {\bibinfo {author} {\bibfnamefont {Abhishek~V.}\ \bibnamefont {Joshi}}, \bibinfo {author} {\bibfnamefont {Shawn~G.}\ \bibnamefont {Rosofsky}}, \bibinfo {author} {\bibfnamefont {Roland}\ \bibnamefont {Haas}}, \ and\ \bibinfo {author} {\bibfnamefont {E.~A.}\ \bibnamefont {Huerta}},\ }\bibfield  {title} {\enquote {\bibinfo {title} {{Numerical relativity higher order gravitational waveforms of eccentric, spinning, nonprecessing binary black hole mergers}},}\ }\href {\doibase 10.1103/PhysRevD.107.064038} {\bibfield  {journal} {\bibinfo  {journal} {Phys. Rev. D}\ }\textbf {\bibinfo {volume} {107}},\ \bibinfo {pages} {064038} (\bibinfo {year} {2023})},\ \Eprint {http://arxiv.org/abs/2210.01852} {arXiv:2210.01852 [gr-qc]} \BibitemShut {NoStop}%
\bibitem [{\citenamefont {Setyawati}\ and\ \citenamefont {Ohme}(2021)}]{Setyawati:2021gom}%
  \BibitemOpen
  \bibfield  {author} {\bibinfo {author} {\bibfnamefont {Yoshinta}\ \bibnamefont {Setyawati}}\ and\ \bibinfo {author} {\bibfnamefont {Frank}\ \bibnamefont {Ohme}},\ }\bibfield  {title} {\enquote {\bibinfo {title} {{Adding eccentricity to quasicircular binary-black-hole waveform models}},}\ }\href {\doibase 10.1103/PhysRevD.103.124011} {\bibfield  {journal} {\bibinfo  {journal} {Phys. Rev. D}\ }\textbf {\bibinfo {volume} {103}},\ \bibinfo {pages} {124011} (\bibinfo {year} {2021})},\ \Eprint {http://arxiv.org/abs/2101.11033} {arXiv:2101.11033 [gr-qc]} \BibitemShut {NoStop}%
\bibitem [{\citenamefont {Wang}\ \emph {et~al.}(2023)\citenamefont {Wang}, \citenamefont {Zou},\ and\ \citenamefont {Liu}}]{Wang:2023ueg}%
  \BibitemOpen
  \bibfield  {author} {\bibinfo {author} {\bibfnamefont {Hao}\ \bibnamefont {Wang}}, \bibinfo {author} {\bibfnamefont {Yuan-Chuan}\ \bibnamefont {Zou}}, \ and\ \bibinfo {author} {\bibfnamefont {Yu}~\bibnamefont {Liu}},\ }\bibfield  {title} {\enquote {\bibinfo {title} {{Phenomenological relationship between eccentric and quasicircular orbital binary black hole waveform}},}\ }\href {\doibase 10.1103/PhysRevD.107.124061} {\bibfield  {journal} {\bibinfo  {journal} {Phys. Rev. D}\ }\textbf {\bibinfo {volume} {107}},\ \bibinfo {pages} {124061} (\bibinfo {year} {2023})},\ \Eprint {http://arxiv.org/abs/2302.11227} {arXiv:2302.11227 [gr-qc]} \BibitemShut {NoStop}%
\bibitem [{\citenamefont {Islam}\ \emph {et~al.}(2021)\citenamefont {Islam}, \citenamefont {Varma}, \citenamefont {Lodman}, \citenamefont {Field}, \citenamefont {Khanna}, \citenamefont {Scheel}, \citenamefont {Pfeiffer}, \citenamefont {Gerosa},\ and\ \citenamefont {Kidder}}]{Islam:2021mha}%
  \BibitemOpen
  \bibfield  {author} {\bibinfo {author} {\bibfnamefont {Tousif}\ \bibnamefont {Islam}}, \bibinfo {author} {\bibfnamefont {Vijay}\ \bibnamefont {Varma}}, \bibinfo {author} {\bibfnamefont {Jackie}\ \bibnamefont {Lodman}}, \bibinfo {author} {\bibfnamefont {Scott~E.}\ \bibnamefont {Field}}, \bibinfo {author} {\bibfnamefont {Gaurav}\ \bibnamefont {Khanna}}, \bibinfo {author} {\bibfnamefont {Mark~A.}\ \bibnamefont {Scheel}}, \bibinfo {author} {\bibfnamefont {Harald~P.}\ \bibnamefont {Pfeiffer}}, \bibinfo {author} {\bibfnamefont {Davide}\ \bibnamefont {Gerosa}}, \ and\ \bibinfo {author} {\bibfnamefont {Lawrence~E.}\ \bibnamefont {Kidder}},\ }\bibfield  {title} {\enquote {\bibinfo {title} {{Eccentric binary black hole surrogate models for the gravitational waveform and remnant properties: comparable mass, nonspinning case}},}\ }\href {\doibase 10.1103/PhysRevD.103.064022} {\bibfield  {journal} {\bibinfo  {journal} {Phys. Rev. D}\ }\textbf {\bibinfo {volume} {103}},\ \bibinfo {pages} {064022} (\bibinfo {year}
  {2021})},\ \Eprint {http://arxiv.org/abs/2101.11798} {arXiv:2101.11798 [gr-qc]} \BibitemShut {NoStop}%
\bibitem [{\citenamefont {Carullo}\ \emph {et~al.}(2024)\citenamefont {Carullo}, \citenamefont {Albanesi}, \citenamefont {Nagar}, \citenamefont {Gamba}, \citenamefont {Bernuzzi}, \citenamefont {Andrade},\ and\ \citenamefont {Trenado}}]{Carullo:2023kvj}%
  \BibitemOpen
  \bibfield  {author} {\bibinfo {author} {\bibfnamefont {Gregorio}\ \bibnamefont {Carullo}}, \bibinfo {author} {\bibfnamefont {Simone}\ \bibnamefont {Albanesi}}, \bibinfo {author} {\bibfnamefont {Alessandro}\ \bibnamefont {Nagar}}, \bibinfo {author} {\bibfnamefont {Rossella}\ \bibnamefont {Gamba}}, \bibinfo {author} {\bibfnamefont {Sebastiano}\ \bibnamefont {Bernuzzi}}, \bibinfo {author} {\bibfnamefont {Tomas}\ \bibnamefont {Andrade}}, \ and\ \bibinfo {author} {\bibfnamefont {Juan}\ \bibnamefont {Trenado}},\ }\bibfield  {title} {\enquote {\bibinfo {title} {{Unveiling the Merger Structure of Black Hole Binaries in Generic Planar Orbits}},}\ }\href {\doibase 10.1103/PhysRevLett.132.101401} {\bibfield  {journal} {\bibinfo  {journal} {Phys. Rev. Lett.}\ }\textbf {\bibinfo {volume} {132}},\ \bibinfo {pages} {101401} (\bibinfo {year} {2024})},\ \Eprint {http://arxiv.org/abs/2309.07228} {arXiv:2309.07228 [gr-qc]} \BibitemShut {NoStop}%
\bibitem [{\citenamefont {Nagar}\ \emph {et~al.}(2021)\citenamefont {Nagar}, \citenamefont {Bonino},\ and\ \citenamefont {Rettegno}}]{Nagar:2021gss}%
  \BibitemOpen
  \bibfield  {author} {\bibinfo {author} {\bibfnamefont {Alessandro}\ \bibnamefont {Nagar}}, \bibinfo {author} {\bibfnamefont {Alice}\ \bibnamefont {Bonino}}, \ and\ \bibinfo {author} {\bibfnamefont {Piero}\ \bibnamefont {Rettegno}},\ }\bibfield  {title} {\enquote {\bibinfo {title} {{Effective one-body multipolar waveform model for spin-aligned, quasicircular, eccentric, hyperbolic black hole binaries}},}\ }\href {\doibase 10.1103/PhysRevD.103.104021} {\bibfield  {journal} {\bibinfo  {journal} {Phys. Rev. D}\ }\textbf {\bibinfo {volume} {103}},\ \bibinfo {pages} {104021} (\bibinfo {year} {2021})},\ \Eprint {http://arxiv.org/abs/2101.08624} {arXiv:2101.08624 [gr-qc]} \BibitemShut {NoStop}%
\bibitem [{\citenamefont {Paul}\ \emph {et~al.}(2024)\citenamefont {Paul}, \citenamefont {Maurya}, \citenamefont {Henry}, \citenamefont {Sharma}, \citenamefont {Satheesh}, \citenamefont {Divyajyoti}, \citenamefont {Kumar},\ and\ \citenamefont {Mishra}}]{Paul:2024ujx}%
  \BibitemOpen
  \bibfield  {author} {\bibinfo {author} {\bibfnamefont {Kaushik}\ \bibnamefont {Paul}}, \bibinfo {author} {\bibfnamefont {Akash}\ \bibnamefont {Maurya}}, \bibinfo {author} {\bibfnamefont {Quentin}\ \bibnamefont {Henry}}, \bibinfo {author} {\bibfnamefont {Kartikey}\ \bibnamefont {Sharma}}, \bibinfo {author} {\bibfnamefont {Pranav}\ \bibnamefont {Satheesh}}, \bibinfo {author} {\bibnamefont {Divyajyoti}}, \bibinfo {author} {\bibfnamefont {Prayush}\ \bibnamefont {Kumar}}, \ and\ \bibinfo {author} {\bibfnamefont {Chandra~Kant}\ \bibnamefont {Mishra}},\ }\bibfield  {title} {\enquote {\bibinfo {title} {{ESIGMAHM: An Eccentric, Spinning inspiral-merger-ringdown waveform model with Higher Modes for the detection and characterization of binary black holes}},}\ }\href@noop {} {\  (\bibinfo {year} {2024})},\ \Eprint {http://arxiv.org/abs/2409.13866} {arXiv:2409.13866 [gr-qc]} \BibitemShut {NoStop}%
\bibitem [{\citenamefont {Manna}\ \emph {et~al.}(2024)\citenamefont {Manna}, \citenamefont {RoyChowdhury},\ and\ \citenamefont {Mishra}}]{Manna:2024ycx}%
  \BibitemOpen
  \bibfield  {author} {\bibinfo {author} {\bibfnamefont {Pratul}\ \bibnamefont {Manna}}, \bibinfo {author} {\bibfnamefont {Tamal}\ \bibnamefont {RoyChowdhury}}, \ and\ \bibinfo {author} {\bibfnamefont {Chandra~Kant}\ \bibnamefont {Mishra}},\ }\bibfield  {title} {\enquote {\bibinfo {title} {{An improved IMR model for BBHs on elliptical orbits}},}\ }\href@noop {} {\  (\bibinfo {year} {2024})},\ \Eprint {http://arxiv.org/abs/2409.10672} {arXiv:2409.10672 [gr-qc]} \BibitemShut {NoStop}%
\bibitem [{\citenamefont {Gamboa}\ \emph {et~al.}(2024{\natexlab{a}})\citenamefont {Gamboa} \emph {et~al.}}]{Gamboa:2024hli}%
  \BibitemOpen
  \bibfield  {author} {\bibinfo {author} {\bibfnamefont {Aldo}\ \bibnamefont {Gamboa}} \emph {et~al.},\ }\bibfield  {title} {\enquote {\bibinfo {title} {{Accurate waveforms for eccentric, aligned-spin binary black holes: The multipolar effective-one-body model SEOBNRv5EHM}},}\ }\href@noop {} {\  (\bibinfo {year} {2024}{\natexlab{a}})},\ \Eprint {http://arxiv.org/abs/2412.12823} {arXiv:2412.12823 [gr-qc]} \BibitemShut {NoStop}%
\bibitem [{\citenamefont {Gamboa}\ \emph {et~al.}(2024{\natexlab{b}})\citenamefont {Gamboa}, \citenamefont {Khalil},\ and\ \citenamefont {Buonanno}}]{Gamboa:2024imd}%
  \BibitemOpen
  \bibfield  {author} {\bibinfo {author} {\bibfnamefont {Aldo}\ \bibnamefont {Gamboa}}, \bibinfo {author} {\bibfnamefont {Mohammed}\ \bibnamefont {Khalil}}, \ and\ \bibinfo {author} {\bibfnamefont {Alessandra}\ \bibnamefont {Buonanno}},\ }\bibfield  {title} {\enquote {\bibinfo {title} {{Third post-Newtonian dynamics for eccentric orbits and aligned spins in the effective-one-body waveform model SEOBNRv5EHM}},}\ }\href@noop {} {\  (\bibinfo {year} {2024}{\natexlab{b}})},\ \Eprint {http://arxiv.org/abs/2412.12831} {arXiv:2412.12831 [gr-qc]} \BibitemShut {NoStop}%
\bibitem [{\citenamefont {Buonanno}\ and\ \citenamefont {Damour}(1999)}]{Buonanno:1998gg}%
  \BibitemOpen
  \bibfield  {author} {\bibinfo {author} {\bibfnamefont {A.}~\bibnamefont {Buonanno}}\ and\ \bibinfo {author} {\bibfnamefont {T.}~\bibnamefont {Damour}},\ }\bibfield  {title} {\enquote {\bibinfo {title} {{Effective one-body approach to general relativistic two-body dynamics}},}\ }\href {\doibase 10.1103/PhysRevD.59.084006} {\bibfield  {journal} {\bibinfo  {journal} {Phys. Rev. D}\ }\textbf {\bibinfo {volume} {59}},\ \bibinfo {pages} {084006} (\bibinfo {year} {1999})},\ \Eprint {http://arxiv.org/abs/gr-qc/9811091} {arXiv:gr-qc/9811091} \BibitemShut {NoStop}%
\bibitem [{\citenamefont {Buonanno}\ and\ \citenamefont {Damour}(2000)}]{Buonanno:2000ef}%
  \BibitemOpen
  \bibfield  {author} {\bibinfo {author} {\bibfnamefont {Alessandra}\ \bibnamefont {Buonanno}}\ and\ \bibinfo {author} {\bibfnamefont {Thibault}\ \bibnamefont {Damour}},\ }\bibfield  {title} {\enquote {\bibinfo {title} {{Transition from inspiral to plunge in binary black hole coalescences}},}\ }\href {\doibase 10.1103/PhysRevD.62.064015} {\bibfield  {journal} {\bibinfo  {journal} {Phys. Rev. D}\ }\textbf {\bibinfo {volume} {62}},\ \bibinfo {pages} {064015} (\bibinfo {year} {2000})},\ \Eprint {http://arxiv.org/abs/gr-qc/0001013} {arXiv:gr-qc/0001013} \BibitemShut {NoStop}%
\bibitem [{\citenamefont {Romero-Shaw}\ \emph {et~al.}(2020)\citenamefont {Romero-Shaw}, \citenamefont {Lasky}, \citenamefont {Thrane},\ and\ \citenamefont {Bustillo}}]{Romero-Shaw:2020thy}%
  \BibitemOpen
  \bibfield  {author} {\bibinfo {author} {\bibfnamefont {Isobel~M.}\ \bibnamefont {Romero-Shaw}}, \bibinfo {author} {\bibfnamefont {Paul~D.}\ \bibnamefont {Lasky}}, \bibinfo {author} {\bibfnamefont {Eric}\ \bibnamefont {Thrane}}, \ and\ \bibinfo {author} {\bibfnamefont {Juan~Calderon}\ \bibnamefont {Bustillo}},\ }\bibfield  {title} {\enquote {\bibinfo {title} {{GW190521: orbital eccentricity and signatures of dynamical formation in a binary black hole merger signal}},}\ }\href {\doibase 10.3847/2041-8213/abbe26} {\bibfield  {journal} {\bibinfo  {journal} {Astrophys. J. Lett.}\ }\textbf {\bibinfo {volume} {903}},\ \bibinfo {pages} {L5} (\bibinfo {year} {2020})},\ \Eprint {http://arxiv.org/abs/2009.04771} {arXiv:2009.04771 [astro-ph.HE]} \BibitemShut {NoStop}%
\bibitem [{\citenamefont {Gayathri}\ \emph {et~al.}(2022)\citenamefont {Gayathri}, \citenamefont {Healy}, \citenamefont {Lange}, \citenamefont {O'Brien}, \citenamefont {Szczepanczyk}, \citenamefont {Bartos}, \citenamefont {Campanelli}, \citenamefont {Klimenko}, \citenamefont {Lousto},\ and\ \citenamefont {O'Shaughnessy}}]{Gayathri:2020coq}%
  \BibitemOpen
  \bibfield  {author} {\bibinfo {author} {\bibfnamefont {V.}~\bibnamefont {Gayathri}}, \bibinfo {author} {\bibfnamefont {J.}~\bibnamefont {Healy}}, \bibinfo {author} {\bibfnamefont {J.}~\bibnamefont {Lange}}, \bibinfo {author} {\bibfnamefont {B.}~\bibnamefont {O'Brien}}, \bibinfo {author} {\bibfnamefont {M.}~\bibnamefont {Szczepanczyk}}, \bibinfo {author} {\bibfnamefont {Imre}\ \bibnamefont {Bartos}}, \bibinfo {author} {\bibfnamefont {M.}~\bibnamefont {Campanelli}}, \bibinfo {author} {\bibfnamefont {S.}~\bibnamefont {Klimenko}}, \bibinfo {author} {\bibfnamefont {C.~O.}\ \bibnamefont {Lousto}}, \ and\ \bibinfo {author} {\bibfnamefont {R.}~\bibnamefont {O'Shaughnessy}},\ }\bibfield  {title} {\enquote {\bibinfo {title} {{Eccentricity estimate for black hole mergers with numerical relativity simulations}},}\ }\href {\doibase 10.1038/s41550-021-01568-w} {\bibfield  {journal} {\bibinfo  {journal} {Nature Astron.}\ }\textbf {\bibinfo {volume} {6}},\ \bibinfo {pages} {344--349} (\bibinfo {year} {2022})},\ \Eprint
  {http://arxiv.org/abs/2009.05461} {arXiv:2009.05461 [astro-ph.HE]} \BibitemShut {NoStop}%
\bibitem [{\citenamefont {Gamba}\ \emph {et~al.}(2023)\citenamefont {Gamba}, \citenamefont {Breschi}, \citenamefont {Carullo}, \citenamefont {Albanesi}, \citenamefont {Rettegno}, \citenamefont {Bernuzzi},\ and\ \citenamefont {Nagar}}]{Gamba:2021gap}%
  \BibitemOpen
  \bibfield  {author} {\bibinfo {author} {\bibfnamefont {Rossella}\ \bibnamefont {Gamba}}, \bibinfo {author} {\bibfnamefont {Matteo}\ \bibnamefont {Breschi}}, \bibinfo {author} {\bibfnamefont {Gregorio}\ \bibnamefont {Carullo}}, \bibinfo {author} {\bibfnamefont {Simone}\ \bibnamefont {Albanesi}}, \bibinfo {author} {\bibfnamefont {Piero}\ \bibnamefont {Rettegno}}, \bibinfo {author} {\bibfnamefont {Sebastiano}\ \bibnamefont {Bernuzzi}}, \ and\ \bibinfo {author} {\bibfnamefont {Alessandro}\ \bibnamefont {Nagar}},\ }\bibfield  {title} {\enquote {\bibinfo {title} {{GW190521 as a dynamical capture of two nonspinning black holes}},}\ }\href {\doibase 10.1038/s41550-022-01813-w} {\bibfield  {journal} {\bibinfo  {journal} {Nature Astron.}\ }\textbf {\bibinfo {volume} {7}},\ \bibinfo {pages} {11--17} (\bibinfo {year} {2023})},\ \Eprint {http://arxiv.org/abs/2106.05575} {arXiv:2106.05575 [gr-qc]} \BibitemShut {NoStop}%
\bibitem [{\citenamefont {Ramos-Buades}\ \emph {et~al.}(2023)\citenamefont {Ramos-Buades}, \citenamefont {Buonanno},\ and\ \citenamefont {Gair}}]{Ramos-Buades:2023yhy}%
  \BibitemOpen
  \bibfield  {author} {\bibinfo {author} {\bibfnamefont {Antoni}\ \bibnamefont {Ramos-Buades}}, \bibinfo {author} {\bibfnamefont {Alessandra}\ \bibnamefont {Buonanno}}, \ and\ \bibinfo {author} {\bibfnamefont {Jonathan}\ \bibnamefont {Gair}},\ }\bibfield  {title} {\enquote {\bibinfo {title} {{Bayesian inference of binary black holes with inspiral-merger-ringdown waveforms using two eccentric parameters}},}\ }\href {\doibase 10.1103/PhysRevD.108.124063} {\bibfield  {journal} {\bibinfo  {journal} {Phys. Rev. D}\ }\textbf {\bibinfo {volume} {108}},\ \bibinfo {pages} {124063} (\bibinfo {year} {2023})},\ \Eprint {http://arxiv.org/abs/2309.15528} {arXiv:2309.15528 [gr-qc]} \BibitemShut {NoStop}%
\bibitem [{\citenamefont {Gupte}\ \emph {et~al.}(2024)\citenamefont {Gupte} \emph {et~al.}}]{Gupte:2024jfe}%
  \BibitemOpen
  \bibfield  {author} {\bibinfo {author} {\bibfnamefont {Nihar}\ \bibnamefont {Gupte}} \emph {et~al.},\ }\bibfield  {title} {\enquote {\bibinfo {title} {{Evidence for eccentricity in the population of binary black holes observed by LIGO-Virgo-KAGRA}},}\ }\href@noop {} {\  (\bibinfo {year} {2024})},\ \Eprint {http://arxiv.org/abs/2404.14286} {arXiv:2404.14286 [gr-qc]} \BibitemShut {NoStop}%
\bibitem [{\citenamefont {Blanchet}(2014)}]{blanchet:2013haa}%
  \BibitemOpen
  \bibfield  {author} {\bibinfo {author} {\bibfnamefont {Luc}\ \bibnamefont {Blanchet}},\ }\bibfield  {title} {\enquote {\bibinfo {title} {{Post-Newtonian Theory for Gravitational Waves}},}\ }\href {\doibase 10.12942/lrr-2014-2} {\bibfield  {journal} {\bibinfo  {journal} {Living Rev. Rel.}\ }\textbf {\bibinfo {volume} {17}},\ \bibinfo {pages} {2} (\bibinfo {year} {2014})},\ \Eprint {http://arxiv.org/abs/1310.1528} {arXiv:1310.1528 [gr-qc]} \BibitemShut {NoStop}%
\bibitem [{\citenamefont {Mora}\ and\ \citenamefont {Will}(2002)}]{Mora:2002gf}%
  \BibitemOpen
  \bibfield  {author} {\bibinfo {author} {\bibfnamefont {Thierry}\ \bibnamefont {Mora}}\ and\ \bibinfo {author} {\bibfnamefont {Clifford~M.}\ \bibnamefont {Will}},\ }\bibfield  {title} {\enquote {\bibinfo {title} {{Numerically generated quasiequilibrium orbits of black holes: Circular or eccentric?}}}\ }\href {\doibase 10.1103/PhysRevD.66.101501} {\bibfield  {journal} {\bibinfo  {journal} {Phys. Rev. D}\ }\textbf {\bibinfo {volume} {66}},\ \bibinfo {pages} {101501} (\bibinfo {year} {2002})},\ \Eprint {http://arxiv.org/abs/gr-qc/0208089} {arXiv:gr-qc/0208089} \BibitemShut {NoStop}%
\bibitem [{\citenamefont {Shaikh}\ \emph {et~al.}(2023)\citenamefont {Shaikh}, \citenamefont {Varma}, \citenamefont {Pfeiffer}, \citenamefont {Ramos-Buades},\ and\ \citenamefont {van~de Meent}}]{Shaikh:2023ypz}%
  \BibitemOpen
  \bibfield  {author} {\bibinfo {author} {\bibfnamefont {Md~Arif}\ \bibnamefont {Shaikh}}, \bibinfo {author} {\bibfnamefont {Vijay}\ \bibnamefont {Varma}}, \bibinfo {author} {\bibfnamefont {Harald~P.}\ \bibnamefont {Pfeiffer}}, \bibinfo {author} {\bibfnamefont {Antoni}\ \bibnamefont {Ramos-Buades}}, \ and\ \bibinfo {author} {\bibfnamefont {Maarten}\ \bibnamefont {van~de Meent}},\ }\bibfield  {title} {\enquote {\bibinfo {title} {{Defining eccentricity for gravitational wave astronomy}},}\ }\href {\doibase 10.1103/PhysRevD.108.104007} {\bibfield  {journal} {\bibinfo  {journal} {Phys. Rev. D}\ }\textbf {\bibinfo {volume} {108}},\ \bibinfo {pages} {104007} (\bibinfo {year} {2023})},\ \Eprint {http://arxiv.org/abs/2302.11257} {arXiv:2302.11257 [gr-qc]} \BibitemShut {NoStop}%
\bibitem [{\citenamefont {Knee}\ \emph {et~al.}(2022)\citenamefont {Knee}, \citenamefont {Romero-Shaw}, \citenamefont {Lasky}, \citenamefont {McIver},\ and\ \citenamefont {Thrane}}]{Knee:2022hth}%
  \BibitemOpen
  \bibfield  {author} {\bibinfo {author} {\bibfnamefont {Alan~M.}\ \bibnamefont {Knee}}, \bibinfo {author} {\bibfnamefont {Isobel~M.}\ \bibnamefont {Romero-Shaw}}, \bibinfo {author} {\bibfnamefont {Paul~D.}\ \bibnamefont {Lasky}}, \bibinfo {author} {\bibfnamefont {Jess}\ \bibnamefont {McIver}}, \ and\ \bibinfo {author} {\bibfnamefont {Eric}\ \bibnamefont {Thrane}},\ }\bibfield  {title} {\enquote {\bibinfo {title} {{A Rosetta Stone for Eccentric Gravitational Waveform Models}},}\ }\href {\doibase 10.3847/1538-4357/ac8b02} {\bibfield  {journal} {\bibinfo  {journal} {Astrophys. J.}\ }\textbf {\bibinfo {volume} {936}},\ \bibinfo {pages} {172} (\bibinfo {year} {2022})},\ \Eprint {http://arxiv.org/abs/2207.14346} {arXiv:2207.14346 [gr-qc]} \BibitemShut {NoStop}%
\bibitem [{\citenamefont {Boschini}\ \emph {et~al.}(2025)\citenamefont {Boschini}, \citenamefont {Loutrel}, \citenamefont {Gerosa},\ and\ \citenamefont {Fumagalli}}]{Boschini:2024scu}%
  \BibitemOpen
  \bibfield  {author} {\bibinfo {author} {\bibfnamefont {Matteo}\ \bibnamefont {Boschini}}, \bibinfo {author} {\bibfnamefont {Nicholas}\ \bibnamefont {Loutrel}}, \bibinfo {author} {\bibfnamefont {Davide}\ \bibnamefont {Gerosa}}, \ and\ \bibinfo {author} {\bibfnamefont {Giulia}\ \bibnamefont {Fumagalli}},\ }\bibfield  {title} {\enquote {\bibinfo {title} {{Orbital eccentricity in general relativity from catastrophe theory}},}\ }\href {\doibase 10.1103/PhysRevD.111.024008} {\bibfield  {journal} {\bibinfo  {journal} {Phys. Rev. D}\ }\textbf {\bibinfo {volume} {111}},\ \bibinfo {pages} {024008} (\bibinfo {year} {2025})},\ \Eprint {http://arxiv.org/abs/2411.00098} {arXiv:2411.00098 [gr-qc]} \BibitemShut {NoStop}%
\bibitem [{\citenamefont {Islam}\ and\ \citenamefont {Venumadhav}(2025)}]{Islam:2025oiv}%
  \BibitemOpen
  \bibfield  {author} {\bibinfo {author} {\bibfnamefont {Tousif}\ \bibnamefont {Islam}}\ and\ \bibinfo {author} {\bibfnamefont {Tejaswi}\ \bibnamefont {Venumadhav}},\ }\bibfield  {title} {\enquote {\bibinfo {title} {{Post-Newtonian theory-inspired framework for characterizing eccentricity in gravitational waveforms}},}\ }\href@noop {} {\  (\bibinfo {year} {2025})},\ \Eprint {http://arxiv.org/abs/2502.02739} {arXiv:2502.02739 [gr-qc]} \BibitemShut {NoStop}%
\bibitem [{\citenamefont {Yunes}\ \emph {et~al.}(2009)\citenamefont {Yunes}, \citenamefont {Arun}, \citenamefont {Berti},\ and\ \citenamefont {Will}}]{Yunes:2009yz}%
  \BibitemOpen
  \bibfield  {author} {\bibinfo {author} {\bibfnamefont {Nicolas}\ \bibnamefont {Yunes}}, \bibinfo {author} {\bibfnamefont {K.~G.}\ \bibnamefont {Arun}}, \bibinfo {author} {\bibfnamefont {Emanuele}\ \bibnamefont {Berti}}, \ and\ \bibinfo {author} {\bibfnamefont {Clifford~M.}\ \bibnamefont {Will}},\ }\bibfield  {title} {\enquote {\bibinfo {title} {{Post-Circular Expansion of Eccentric Binary Inspirals: Fourier-Domain Waveforms in the Stationary Phase Approximation}},}\ }\href {\doibase 10.1103/PhysRevD.80.084001} {\bibfield  {journal} {\bibinfo  {journal} {Phys. Rev. D}\ }\textbf {\bibinfo {volume} {80}},\ \bibinfo {pages} {084001} (\bibinfo {year} {2009})},\ \bibinfo {note} {[Erratum: Phys.Rev.D 89, 109901 (2014)]},\ \Eprint {http://arxiv.org/abs/0906.0313} {arXiv:0906.0313 [gr-qc]} \BibitemShut {NoStop}%
\bibitem [{\citenamefont {Van Den~Broeck}\ and\ \citenamefont {Sengupta}(2007)}]{VanDenBroeck:2006qu}%
  \BibitemOpen
  \bibfield  {author} {\bibinfo {author} {\bibfnamefont {Chris}\ \bibnamefont {Van Den~Broeck}}\ and\ \bibinfo {author} {\bibfnamefont {Anand~S.}\ \bibnamefont {Sengupta}},\ }\bibfield  {title} {\enquote {\bibinfo {title} {{Phenomenology of amplitude-corrected post-Newtonian gravitational waveforms for compact binary inspiral. I. Signal-to-noise ratios}},}\ }\href {\doibase 10.1088/0264-9381/24/1/009} {\bibfield  {journal} {\bibinfo  {journal} {Class. Quant. Grav.}\ }\textbf {\bibinfo {volume} {24}},\ \bibinfo {pages} {155--176} (\bibinfo {year} {2007})},\ \Eprint {http://arxiv.org/abs/gr-qc/0607092} {arXiv:gr-qc/0607092} \BibitemShut {NoStop}%
\bibitem [{\citenamefont {Arun}\ \emph {et~al.}(2007)\citenamefont {Arun}, \citenamefont {Iyer}, \citenamefont {Sathyaprakash},\ and\ \citenamefont {Sinha}}]{Arun:2007qv}%
  \BibitemOpen
  \bibfield  {author} {\bibinfo {author} {\bibfnamefont {K.~G.}\ \bibnamefont {Arun}}, \bibinfo {author} {\bibfnamefont {Bala~R.}\ \bibnamefont {Iyer}}, \bibinfo {author} {\bibfnamefont {B.~S.}\ \bibnamefont {Sathyaprakash}}, \ and\ \bibinfo {author} {\bibfnamefont {Siddhartha}\ \bibnamefont {Sinha}},\ }\bibfield  {title} {\enquote {\bibinfo {title} {{Higher harmonics increase LISA's mass reach for supermassive black holes}},}\ }\href {\doibase 10.1103/PhysRevD.75.124002} {\bibfield  {journal} {\bibinfo  {journal} {Phys. Rev. D}\ }\textbf {\bibinfo {volume} {75}},\ \bibinfo {pages} {124002} (\bibinfo {year} {2007})},\ \Eprint {http://arxiv.org/abs/0704.1086} {arXiv:0704.1086 [gr-qc]} \BibitemShut {NoStop}%
\bibitem [{\citenamefont {Seto}(2001)}]{Seto:2001pg}%
  \BibitemOpen
  \bibfield  {author} {\bibinfo {author} {\bibfnamefont {Naoki}\ \bibnamefont {Seto}},\ }\bibfield  {title} {\enquote {\bibinfo {title} {{Proposal for determining the total masses of eccentric binaries using signature of periastron advance in gravitational waves}},}\ }\href {\doibase 10.1103/PhysRevLett.87.251101} {\bibfield  {journal} {\bibinfo  {journal} {Phys. Rev. Lett.}\ }\textbf {\bibinfo {volume} {87}},\ \bibinfo {pages} {251101} (\bibinfo {year} {2001})},\ \bibinfo {note} {[Erratum: Phys.Rev.Lett. 101, 209901 (2008)]},\ \Eprint {http://arxiv.org/abs/astro-ph/0111107} {arXiv:astro-ph/0111107} \BibitemShut {NoStop}%
\bibitem [{\citenamefont {{Valsecchi}}\ \emph {et~al.}(2012)\citenamefont {{Valsecchi}}, \citenamefont {{Farr}}, \citenamefont {{Willems}}, \citenamefont {{Deloye}},\ and\ \citenamefont {{Kalogera}}}]{2012ApJ74537V}%
  \BibitemOpen
  \bibfield  {author} {\bibinfo {author} {\bibfnamefont {F.}~\bibnamefont {{Valsecchi}}}, \bibinfo {author} {\bibfnamefont {W.~M.}\ \bibnamefont {{Farr}}}, \bibinfo {author} {\bibfnamefont {B.}~\bibnamefont {{Willems}}}, \bibinfo {author} {\bibfnamefont {C.~J.}\ \bibnamefont {{Deloye}}}, \ and\ \bibinfo {author} {\bibfnamefont {V.}~\bibnamefont {{Kalogera}}},\ }\bibfield  {title} {\enquote {\bibinfo {title} {{Tidally Induced Apsidal Precession in Double White Dwarfs: A New Mass Measurement Tool with LISA}},}\ }\href {\doibase 10.1088/0004-637X/745/2/137} {\bibfield  {journal} {\bibinfo  {journal} {\apj}\ }\textbf {\bibinfo {volume} {745}},\ \bibinfo {eid} {137} (\bibinfo {year} {2012})},\ \Eprint {http://arxiv.org/abs/1105.4837} {arXiv:1105.4837 [astro-ph.SR]} \BibitemShut {NoStop}%
\bibitem [{\citenamefont {Patterson}\ \emph {et~al.}(2024)\citenamefont {Patterson}, \citenamefont {Tomson},\ and\ \citenamefont {Fairhurst}}]{Patterson:2024vbo}%
  \BibitemOpen
  \bibfield  {author} {\bibinfo {author} {\bibfnamefont {Ben~G.}\ \bibnamefont {Patterson}}, \bibinfo {author} {\bibfnamefont {Sharon~Mary}\ \bibnamefont {Tomson}}, \ and\ \bibinfo {author} {\bibfnamefont {Stephen}\ \bibnamefont {Fairhurst}},\ }\bibfield  {title} {\enquote {\bibinfo {title} {{Identifying Eccentricity in Binary Black Hole mergers using a Harmonic Decomposition of the Gravitational Waveform}},}\ }\href@noop {} {\  (\bibinfo {year} {2024})},\ \Eprint {http://arxiv.org/abs/2411.04187} {arXiv:2411.04187 [gr-qc]} \BibitemShut {NoStop}%
\bibitem [{\citenamefont {Islam}\ \emph {et~al.}(2025)\citenamefont {Islam}, \citenamefont {Venumadhav}, \citenamefont {Mehta}, \citenamefont {Anantpurkar}, \citenamefont {Wadekar}, \citenamefont {Roulet}, \citenamefont {Mushkin}, \citenamefont {Zackay},\ and\ \citenamefont {Zaldarriaga}}]{Islam2025InPrep}%
  \BibitemOpen
  \bibfield  {author} {\bibinfo {author} {\bibfnamefont {Tousif}\ \bibnamefont {Islam}}, \bibinfo {author} {\bibfnamefont {Tejaswi}\ \bibnamefont {Venumadhav}}, \bibinfo {author} {\bibfnamefont {Ajit~K.}\ \bibnamefont {Mehta}}, \bibinfo {author} {\bibfnamefont {Isha}\ \bibnamefont {Anantpurkar}}, \bibinfo {author} {\bibfnamefont {Digvijay}\ \bibnamefont {Wadekar}}, \bibinfo {author} {\bibfnamefont {Javier}\ \bibnamefont {Roulet}}, \bibinfo {author} {\bibfnamefont {Jonathan}\ \bibnamefont {Mushkin}}, \bibinfo {author} {\bibfnamefont {Barak}\ \bibnamefont {Zackay}}, \ and\ \bibinfo {author} {\bibfnamefont {Matias}\ \bibnamefont {Zaldarriaga}},\ }\bibfield  {title} {\enquote {\bibinfo {title} {Data-driven extraction, phenomenology and modeling of eccentric harmonics in binary black hole merger waveforms},}\ }\href@noop {} {\  (\bibinfo {year} {2025})},\ \bibinfo {note} {{In preparation}}\BibitemShut {NoStop}%
\bibitem [{\citenamefont {Varma}\ \emph {et~al.}(2019)\citenamefont {Varma}, \citenamefont {Field}, \citenamefont {Scheel}, \citenamefont {Blackman}, \citenamefont {Gerosa}, \citenamefont {Stein}, \citenamefont {Kidder},\ and\ \citenamefont {Pfeiffer}}]{varma2019surrogate}%
  \BibitemOpen
  \bibfield  {author} {\bibinfo {author} {\bibfnamefont {Vijay}\ \bibnamefont {Varma}}, \bibinfo {author} {\bibfnamefont {Scott~E.}\ \bibnamefont {Field}}, \bibinfo {author} {\bibfnamefont {Mark~A.}\ \bibnamefont {Scheel}}, \bibinfo {author} {\bibfnamefont {Jonathan}\ \bibnamefont {Blackman}}, \bibinfo {author} {\bibfnamefont {Davide}\ \bibnamefont {Gerosa}}, \bibinfo {author} {\bibfnamefont {Leo~C.}\ \bibnamefont {Stein}}, \bibinfo {author} {\bibfnamefont {Lawrence~E.}\ \bibnamefont {Kidder}}, \ and\ \bibinfo {author} {\bibfnamefont {Harald~P.}\ \bibnamefont {Pfeiffer}},\ }\bibfield  {title} {\enquote {\bibinfo {title} {{Surrogate models for precessing binary black hole simulations with unequal masses}},}\ }\href {\doibase 10.1103/PhysRevResearch.1.033015} {\bibfield  {journal} {\bibinfo  {journal} {Phys. Rev. Research.}\ }\textbf {\bibinfo {volume} {1}},\ \bibinfo {pages} {033015} (\bibinfo {year} {2019})},\ \Eprint {http://arxiv.org/abs/1905.09300} {arXiv:1905.09300 [gr-qc]} \BibitemShut {NoStop}%
\bibitem [{\citenamefont {Wen}(2003)}]{Wen:2002km}%
  \BibitemOpen
  \bibfield  {author} {\bibinfo {author} {\bibfnamefont {Linqing}\ \bibnamefont {Wen}},\ }\bibfield  {title} {\enquote {\bibinfo {title} {{On the eccentricity distribution of coalescing black hole binaries driven by the Kozai mechanism in globular clusters}},}\ }\href {\doibase 10.1086/378794} {\bibfield  {journal} {\bibinfo  {journal} {Astrophys. J.}\ }\textbf {\bibinfo {volume} {598}},\ \bibinfo {pages} {419--430} (\bibinfo {year} {2003})},\ \Eprint {http://arxiv.org/abs/astro-ph/0211492} {arXiv:astro-ph/0211492} \BibitemShut {NoStop}%
\bibitem [{\citenamefont {Gough}(2009)}]{10.5555/1538674}%
  \BibitemOpen
  \bibfield  {author} {\bibinfo {author} {\bibfnamefont {Brian}\ \bibnamefont {Gough}},\ }\href@noop {} {\emph {\bibinfo {title} {GNU Scientific Library Reference Manual - Third Edition}}},\ \bibinfo {edition} {3rd}\ ed.\ (\bibinfo  {publisher} {Network Theory Ltd.},\ \bibinfo {year} {2009})\BibitemShut {NoStop}%
\bibitem [{\citenamefont {Pathak}\ \emph {et~al.}(2024)\citenamefont {Pathak}, \citenamefont {Reza},\ and\ \citenamefont {Sengupta}}]{Pathak:2024zgo}%
  \BibitemOpen
  \bibfield  {author} {\bibinfo {author} {\bibfnamefont {Lalit}\ \bibnamefont {Pathak}}, \bibinfo {author} {\bibfnamefont {Amit}\ \bibnamefont {Reza}}, \ and\ \bibinfo {author} {\bibfnamefont {Anand~S.}\ \bibnamefont {Sengupta}},\ }\bibfield  {title} {\enquote {\bibinfo {title} {{Fast and faithful interpolation of numerical relativity surrogate waveforms using a meshfree approximation}},}\ }\href {\doibase 10.1103/PhysRevD.110.064022} {\bibfield  {journal} {\bibinfo  {journal} {Phys. Rev. D}\ }\textbf {\bibinfo {volume} {110}},\ \bibinfo {pages} {064022} (\bibinfo {year} {2024})},\ \Eprint {http://arxiv.org/abs/2403.19162} {arXiv:2403.19162 [gr-qc]} \BibitemShut {NoStop}%
\bibitem [{\citenamefont {Wadekar}\ \emph {et~al.}(2024)\citenamefont {Wadekar}, \citenamefont {Venumadhav}, \citenamefont {Roulet}, \citenamefont {Mehta}, \citenamefont {Zackay}, \citenamefont {Mushkin},\ and\ \citenamefont {Zaldarriaga}}]{Wadekar:2024zdq}%
  \BibitemOpen
  \bibfield  {author} {\bibinfo {author} {\bibfnamefont {Digvijay}\ \bibnamefont {Wadekar}}, \bibinfo {author} {\bibfnamefont {Tejaswi}\ \bibnamefont {Venumadhav}}, \bibinfo {author} {\bibfnamefont {Javier}\ \bibnamefont {Roulet}}, \bibinfo {author} {\bibfnamefont {Ajit~Kumar}\ \bibnamefont {Mehta}}, \bibinfo {author} {\bibfnamefont {Barak}\ \bibnamefont {Zackay}}, \bibinfo {author} {\bibfnamefont {Jonathan}\ \bibnamefont {Mushkin}}, \ and\ \bibinfo {author} {\bibfnamefont {Matias}\ \bibnamefont {Zaldarriaga}},\ }\bibfield  {title} {\enquote {\bibinfo {title} {{New search pipeline for gravitational waves with higher-order modes using mode-by-mode filtering}},}\ }\href {\doibase 10.1103/PhysRevD.110.044063} {\bibfield  {journal} {\bibinfo  {journal} {Phys. Rev. D}\ }\textbf {\bibinfo {volume} {110}},\ \bibinfo {pages} {044063} (\bibinfo {year} {2024})},\ \Eprint {http://arxiv.org/abs/2405.17400} {arXiv:2405.17400 [gr-qc]} \BibitemShut {NoStop}%
\bibitem [{\citenamefont {Roulet}\ \emph {et~al.}(2019)\citenamefont {Roulet}, \citenamefont {Dai}, \citenamefont {Venumadhav}, \citenamefont {Zackay},\ and\ \citenamefont {Zaldarriaga}}]{Roulet:2019hzy}%
  \BibitemOpen
  \bibfield  {author} {\bibinfo {author} {\bibfnamefont {Javier}\ \bibnamefont {Roulet}}, \bibinfo {author} {\bibfnamefont {Liang}\ \bibnamefont {Dai}}, \bibinfo {author} {\bibfnamefont {Tejaswi}\ \bibnamefont {Venumadhav}}, \bibinfo {author} {\bibfnamefont {Barak}\ \bibnamefont {Zackay}}, \ and\ \bibinfo {author} {\bibfnamefont {Matias}\ \bibnamefont {Zaldarriaga}},\ }\bibfield  {title} {\enquote {\bibinfo {title} {{Template Bank for Compact Binary Coalescence Searches in Gravitational Wave Data: A General Geometric Placement Algorithm}},}\ }\href {\doibase 10.1103/PhysRevD.99.123022} {\bibfield  {journal} {\bibinfo  {journal} {Phys. Rev. D}\ }\textbf {\bibinfo {volume} {99}},\ \bibinfo {pages} {123022} (\bibinfo {year} {2019})},\ \Eprint {http://arxiv.org/abs/1904.01683} {arXiv:1904.01683 [astro-ph.IM]} \BibitemShut {NoStop}%
\bibitem [{\citenamefont {Blackman}\ \emph {et~al.}(2017)\citenamefont {Blackman}, \citenamefont {Field}, \citenamefont {Scheel}, \citenamefont {Galley}, \citenamefont {Hemberger}, \citenamefont {Schmidt},\ and\ \citenamefont {Smith}}]{Blackman:2017dfb}%
  \BibitemOpen
  \bibfield  {author} {\bibinfo {author} {\bibfnamefont {Jonathan}\ \bibnamefont {Blackman}}, \bibinfo {author} {\bibfnamefont {Scott~E.}\ \bibnamefont {Field}}, \bibinfo {author} {\bibfnamefont {Mark~A.}\ \bibnamefont {Scheel}}, \bibinfo {author} {\bibfnamefont {Chad~R.}\ \bibnamefont {Galley}}, \bibinfo {author} {\bibfnamefont {Daniel~A.}\ \bibnamefont {Hemberger}}, \bibinfo {author} {\bibfnamefont {Patricia}\ \bibnamefont {Schmidt}}, \ and\ \bibinfo {author} {\bibfnamefont {Rory}\ \bibnamefont {Smith}},\ }\bibfield  {title} {\enquote {\bibinfo {title} {{A Surrogate Model of Gravitational Waveforms from Numerical Relativity Simulations of Precessing Binary Black Hole Mergers}},}\ }\href {\doibase 10.1103/PhysRevD.95.104023} {\bibfield  {journal} {\bibinfo  {journal} {Phys. Rev. D}\ }\textbf {\bibinfo {volume} {95}},\ \bibinfo {pages} {104023} (\bibinfo {year} {2017})},\ \Eprint {http://arxiv.org/abs/1701.00550} {arXiv:1701.00550 [gr-qc]} \BibitemShut {NoStop}%
\bibitem [{\citenamefont {Cutler}\ and\ \citenamefont {Flanagan}(1994)}]{Cutler:1994ys}%
  \BibitemOpen
  \bibfield  {author} {\bibinfo {author} {\bibfnamefont {Curt}\ \bibnamefont {Cutler}}\ and\ \bibinfo {author} {\bibfnamefont {Eanna~E.}\ \bibnamefont {Flanagan}},\ }\bibfield  {title} {\enquote {\bibinfo {title} {{Gravitational waves from merging compact binaries: How accurately can one extract the binary's parameters from the inspiral wave form?}}}\ }\href {\doibase 10.1103/PhysRevD.49.2658} {\bibfield  {journal} {\bibinfo  {journal} {Phys. Rev. D}\ }\textbf {\bibinfo {volume} {49}},\ \bibinfo {pages} {2658--2697} (\bibinfo {year} {1994})},\ \Eprint {http://arxiv.org/abs/gr-qc/9402014} {arXiv:gr-qc/9402014} \BibitemShut {NoStop}%
\bibitem [{\citenamefont {Abbott}\ \emph {et~al.}(2016)\citenamefont {Abbott} \emph {et~al.}}]{KAGRA:2013rdx}%
  \BibitemOpen
  \bibfield  {author} {\bibinfo {author} {\bibfnamefont {B.~P.}\ \bibnamefont {Abbott}} \emph {et~al.} (\bibinfo {collaboration} {KAGRA, LIGO Scientific, Virgo}),\ }\bibfield  {title} {\enquote {\bibinfo {title} {{Prospects for observing and localizing gravitational-wave transients with Advanced LIGO, Advanced Virgo and KAGRA}},}\ }\href {\doibase 10.1007/s41114-020-00026-9} {\bibfield  {journal} {\bibinfo  {journal} {Living Rev. Rel.}\ }\textbf {\bibinfo {volume} {19}},\ \bibinfo {pages} {1} (\bibinfo {year} {2016})},\ \Eprint {http://arxiv.org/abs/1304.0670} {arXiv:1304.0670 [gr-qc]} \BibitemShut {NoStop}%
\bibitem [{\citenamefont {Zackay}\ \emph {et~al.}(2018)\citenamefont {Zackay}, \citenamefont {Dai},\ and\ \citenamefont {Venumadhav}}]{Zackay:2018qdy}%
  \BibitemOpen
  \bibfield  {author} {\bibinfo {author} {\bibfnamefont {Barak}\ \bibnamefont {Zackay}}, \bibinfo {author} {\bibfnamefont {Liang}\ \bibnamefont {Dai}}, \ and\ \bibinfo {author} {\bibfnamefont {Tejaswi}\ \bibnamefont {Venumadhav}},\ }\bibfield  {title} {\enquote {\bibinfo {title} {{Relative Binning and Fast Likelihood Evaluation for Gravitational Wave Parameter Estimation}},}\ }\href@noop {} {\  (\bibinfo {year} {2018})},\ \Eprint {http://arxiv.org/abs/1806.08792} {arXiv:1806.08792 [astro-ph.IM]} \BibitemShut {NoStop}%
\bibitem [{\citenamefont {Roulet}\ \emph {et~al.}(2024)\citenamefont {Roulet}, \citenamefont {Mushkin}, \citenamefont {Wadekar}, \citenamefont {Venumadhav}, \citenamefont {Zackay},\ and\ \citenamefont {Zaldarriaga}}]{Roulet:2024hwz}%
  \BibitemOpen
  \bibfield  {author} {\bibinfo {author} {\bibfnamefont {Javier}\ \bibnamefont {Roulet}}, \bibinfo {author} {\bibfnamefont {Jonathan}\ \bibnamefont {Mushkin}}, \bibinfo {author} {\bibfnamefont {Digvijay}\ \bibnamefont {Wadekar}}, \bibinfo {author} {\bibfnamefont {Tejaswi}\ \bibnamefont {Venumadhav}}, \bibinfo {author} {\bibfnamefont {Barak}\ \bibnamefont {Zackay}}, \ and\ \bibinfo {author} {\bibfnamefont {Matias}\ \bibnamefont {Zaldarriaga}},\ }\bibfield  {title} {\enquote {\bibinfo {title} {{Fast marginalization algorithm for optimizing gravitational wave detection, parameter estimation, and sky localization}},}\ }\href {\doibase 10.1103/PhysRevD.110.044010} {\bibfield  {journal} {\bibinfo  {journal} {Phys. Rev. D}\ }\textbf {\bibinfo {volume} {110}},\ \bibinfo {pages} {044010} (\bibinfo {year} {2024})},\ \Eprint {http://arxiv.org/abs/2404.02435} {arXiv:2404.02435 [gr-qc]} \BibitemShut {NoStop}%
\end{thebibliography}%

\end{document}